\documentstyle[12pt,aaspp4,epsf]{article}
\newcommand\ltorder{\mathrel{\raise.3ex\hbox{$<$}\mkern-14mu
             \lower0.6ex\hbox{$\sim$}}}
\newcommand\gtorder{\mathrel{\raise.3ex\hbox{$>$}\mkern-14mu
             \lower0.6ex\hbox{$\sim$}}}

\begin{document}

\title{Structure of Dark Matter Halos From Hierarchical Clustering}

\author{Toshiyuki Fukushige}

\affil{Department of General Systems Studies,\\
College of Arts and Sciences, University of Tokyo,\\
 3-8-1 Komaba, Meguro-ku, Tokyo 153, Japan}

\author{Junichiro Makino}

\affil{Department of Astronomy,\\
School of Sciences, University of Tokyo,\\
 7-3-1 Hongo, Bunkyo-ku, Tokyo 117, Japan}

%\centerline{\bf (Version of \today)}

\begin{abstract}

We investigate the structure of the dark matter halo formed in the cold
dark matter scenario using $N$-body simulations.  We simulated 12 halos
with the mass of $6.6\times 10^{11}M_{\odot}$ to $8.0\times
10^{14}M_{\odot}$.  In almost all runs, the halos have density cusps
proportional to $r^{-1.5}$ developed at the center, which is consistent
with the results of recent high-resolution calculations.  The density
structure evolves in a self-similar way, and is universal in the sense
that it is independent of the halo mass and initial random realization
of density fluctuation.  The density profile is in good agreement with
the profile proposed by Moore et al.  (1999), which has central slope
proportional to $r^{-1.5}$ and outer slope proportional to $r^{-3}$. 
The halo grows through repeated accretion of diffuse smaller halos.  We
argue that the cusp is understood as a convergence slope for the
accretion of tidally disrupted matter. 

\end{abstract}

\keywords{cosmology:theory --- dark matter --- galaxies:kinematics and
dynamics --- galaxies:formation --- method: numerical}

\section{Introduction}

In standard cosmological pictures, such as the cold dark matter
cosmology, dark matter halos are considered to be formed in a
hierarchical way; smaller halos first formed from initial density
fluctuations and they merged with each other to become larger halos.  In
reality, the formation process of dark matter halo is rather
complicated, since a variety of processes, such as merging between halos
of various sizes and tidal disruption of small halos (satellite) proceed
simultaneously. 

One of the most influential works on the dark matter halo is the
"finding" of the universal profile by Navarro, Frenk, and White (1996,
1997, hereafter NFW), though there were many analytical and numerical
studies before NFW (see NFW (1996) or Bertschinger (1998) for reviews).  
NFW performed $N$-body simulations of the halo formation and
found that the profile of dark matter halo can be fitted by a simple formula
\begin{equation} 
\rho = {\rho_0 \over (r/r_{\rm s})(1+r/r_{\rm s})^2},
\end{equation} 
where $\rho_0$ is a characteristic density and $r_{\rm s}$ is a scale
radius.  They also argued that the profile has the same shape,
independent of the halo mass, the initial density fluctuation spectrum
or the value of the cosmological parameters.  It should be noted that,
before NFW, Dubinski and Carlberg (1991) also found in their
high-resolution simulation the halo can be well fitted by Hernquist
(1990) profile . 

Many studies on the NFW "universal profile", both numerical and
analytical, were done after their proposal.  Many $N$-body simulations
whose resolution are similar to those of NFW were performed and results
similar to NFW were obtained (Cole and Lacy 1996, Tormen, Bouchet and
White 1996, Brainerd, Goldberg, and Villumsen 1998, Thomas et al.  1998,
Okamoto and Habe 1999, Huss, Jain and Steinmetz 1999, Kravtov et al. 
1998, Jing 2000).  Analytical and semi-analytical studies to explain the
NFW universal profile were also done (Evans and Collet 1997, Syer and
White 1998, Avila-Reese, Firmani, Hernandez 1998, Nusser and Sheth 1999,
Kull 1999, Heriksen and Widrow 1999, Yano and Gouda 1999, Bullock et al. 
1999, Subramanian, Cen, Ostriker 2000, Lokas 2000).  The clear
understanding for the NFW profile, however, has not yet been given.  One
of the reasons why a clear understanding has not been established might
be that all these studies were trying to answer a wrong question. 

Our previous study (Fukushige and Makino 1997, hereafter FM97) showed
that density profile obtained by high-resolution $N$-body simulation is
different from the NFW universal profile.  We performed simulations with
768k particles, while previous studies employed $\sim 20$k.  We found
that the galaxy-sized halo has a cusp steeper than $\rho\propto r^{-1}$. 

This disagreement with the NFW universal profile was confirmed by other
high-resolution simulations.  Moore et al.  (1998, 1999) and Ghigna et
al.  (2000) performed simulations with up to 4M particles and obtained
the results similar to ours.  They found that cluster-sized halos also
have cusps steeper than the NFW profile and they proposed the modified
universal profile, $\rho = \rho_0 / [(r/r_{\rm s})^{1.5}(1+(r/r_{\rm
s})^{1.5})]$.  On the other hand, Jing and Suto (2000) found that the
density profile of dark matter is not universal.  They performed a
series of $N$-body simulations and concluded that the power of the cusp
depends on mass.  It varies from $-1.5$ for galaxy mass halo to $-1.1$
for cluster mass halo. 

In this paper, we again investigate the structure of dark matter halos
using $N$-body simulation.  We performed $N$-body simulations of
formation of 12 dark matter halos with masses $6.6\times
10^{11}M_{\odot}$ to $8.0\times 10^{14}M_{\odot}$, using a
special-purpose computer GRAPE-5 (Kawai et al.  2000) and Barnes-Hut
treecode.  In section 2, we describe the model of our $N$-body
simulation.  In section 3, we present the results of simulation. 
Section 4 is for summary and section 5 is for discussion. 

\section{Simulations Models}

We performed in total 12 runs on 4 different mass scales, which are
summarized in Table 1.  Initial conditions were constructed in a way
similar to that in FM97.  We assigned initial positions and velocities
to particles in a spherical region with a radius of $R$ Mpc surrounding
a density peak selected from a unconstrained discrete realization of the
standard CDM model ($H_o = 50$km/s/Mpc, $\Omega = 1$ and
$\sigma_8=0.7$).  The peak was chosen from an $R_{\rm box}$ Mpc cube
using a density field smoothed by a Gaussian filter of radius ($R_{\rm
box}/2$) Mpc.  The values of $R$ and $R_{\rm box}$ in the comoving flame
are summarized in Table 1.  In order to generate the discrete
realization of the CDM model we used the COSMICS package. 

\begin{table}[h]
\caption{Simulation Models}
\begin{tabular}{lccccccc}
\hline
\hline
Run & $R$ (Mpc) & $R_{\rm box}$ (Mpc) & $m$ ($M_{\odot}$)& $\varepsilon$ (kpc) 
& $\Delta t$ (yr) & $z_{\rm start}$ & $z_{\rm end}$ \\
\hline
16M0 & 12.8 & 32  & $3.0\times 10^8$ & 0.56 & $1.6\times 10^6$ & 18.8 & 0.0 \\
16M1 & 16 & 32  & $6.0\times 10^8$ & 0.56 & $1.5\times 10^6$ & 18.5 & 0.0 \\
16M2 & 16 & 32  & $6.1\times 10^8$ & 0.56 & $1.6\times 10^6$ & 20.4 & 0.0 \\
\hline
8M0 & 6.4 & 16  & $3.7\times 10^7$ & 0.28 & $7.9\times 10^5$ & 22.3 & 0.58 \\
8M1 & 8 & 16  & $7.6\times 10^7$ & 0.28 & $7.5\times 10^5$ & 22.2 & 0.63 \\
8M2 & 8 & 16  & $7.6\times 10^7$ & 0.28 & $7.8\times 10^5$ & 23.9 & 0.59 \\
\hline
4M0 & 3.2 & 8  & $4.7\times 10^6$ & 0.14 & $7.8\times 10^5$ & 25.9 & 1.6 \\
4M1 & 4 & 8  & $9.5\times 10^6$ & 0.14 & $7.4\times 10^5$ & 25.9 & 1.6 \\
4M2 & 4 & 8  & $9.5\times 10^6$ & 0.14 & $7.6\times 10^5$ & 27.4 & 1.2 \\
\hline
2M0 & 1.6 & 4  & $5.9\times 10^5$ & 0.07 & $3.8\times 10^5$ & 29.7 & 2.1 \\
2M1 & 2 & 4  & $1.2\times 10^6$ & 0.07 & $3.6\times 10^5$ & 29.7 & 2.2 \\
2M2 & 2 & 4  & $1.2\times 10^6$ & 0.07 & $3.8\times 10^5$ & 30.9 & 1.8 \\
\hline
\end{tabular}
\end{table}

We followed evolution of the density peak by $N$-body simulation.  We
added the local Hubble flow and integrated the orbits directly in the
physical space.  We used the Plummer softened potential with the
softening length constant in physical space, and used a leap-flog
integrator with shared and constant timestep.  In Table 1 we summarized
the individual particle mass, $m$, softening length, $\varepsilon$,
timestep size, $\Delta t$, and starting and ending redshift, $z_{\rm
start}$ and $z_{\rm end}$.  The masses of particles are equal and the
total number of particles for each simulation is $(2.0-2.1) \times
10^6$. 

We determined the radius $R$ Mpc for Run 16M\{0,1,2\} using trial runs
with smaller number of particles, so that all particles lying inside of
$r_{200}$ at $z_{\rm end}$ are included.  Here, the radius $r_{200}$ is
defined as the radius of the sphere in which the mean density $\rho$ is
equal to $200\rho_{\rm crit}$, where $\rho_{\rm crit}$ is the critical
density.  We did not include tidal effects from outside the $R$ Mpc
sphere.  The region of Runs 8M\{0,1,2\}, 4M\{0,1,2\}, and 2M\{0,1,2\} is
1/2, 1/4, and 1/8 of the size of 16M\{0,1,2\} and mass resolution are
$\times 2$, $\times 4$, and $\times 8$, respectively.  The ending
redshift $z_{\rm end}$ for Runs 8M\{0,1,2\}, 4M\{0,1,2\}, and
2M\{0,1,2\} is determined so that the truncation outside the sphere did
not influence the profile around $r_{200}$. 

The number before M in a run name (ex.  8 for Run 8M1) indicates the
length of the simulation box, $R_{\rm box}/2$, in Mpc.  The number after
M in the run name (ex.  1 for Run 8M1) identifies the index for the
random number seed used to generate initial density field.  For example,
Runs \{16,8,4,2\}M0 are series of runs in which the phases of initial
density waves are same and the amplitude of the wave are different. 

For the force calculation, we used the Barnes-Hut tree code
($\theta=0.75$, Barnes and Hut 1986, Barnes 1990, Makino 1991)
implemented on GRAPE-5 (Kawai et al.  2000), a special-purpose computer
designed to accelerate $N$-body simulations.  Using the tree code on two
GRAPE-5 boards and a workstation whose CPU is 21264/677MHz Alpha chip,
one timestep took 21 seconds.  The total number of timesteps is about
$6000-8000$.  Therefore, we can complete one run in 35-50 CPU hours. 
 
\section{Result}

\subsection{Snapshot}

\begin{figure}
\begin{center}
{\leavevmode
\epsfxsize=15cm
\epsfbox{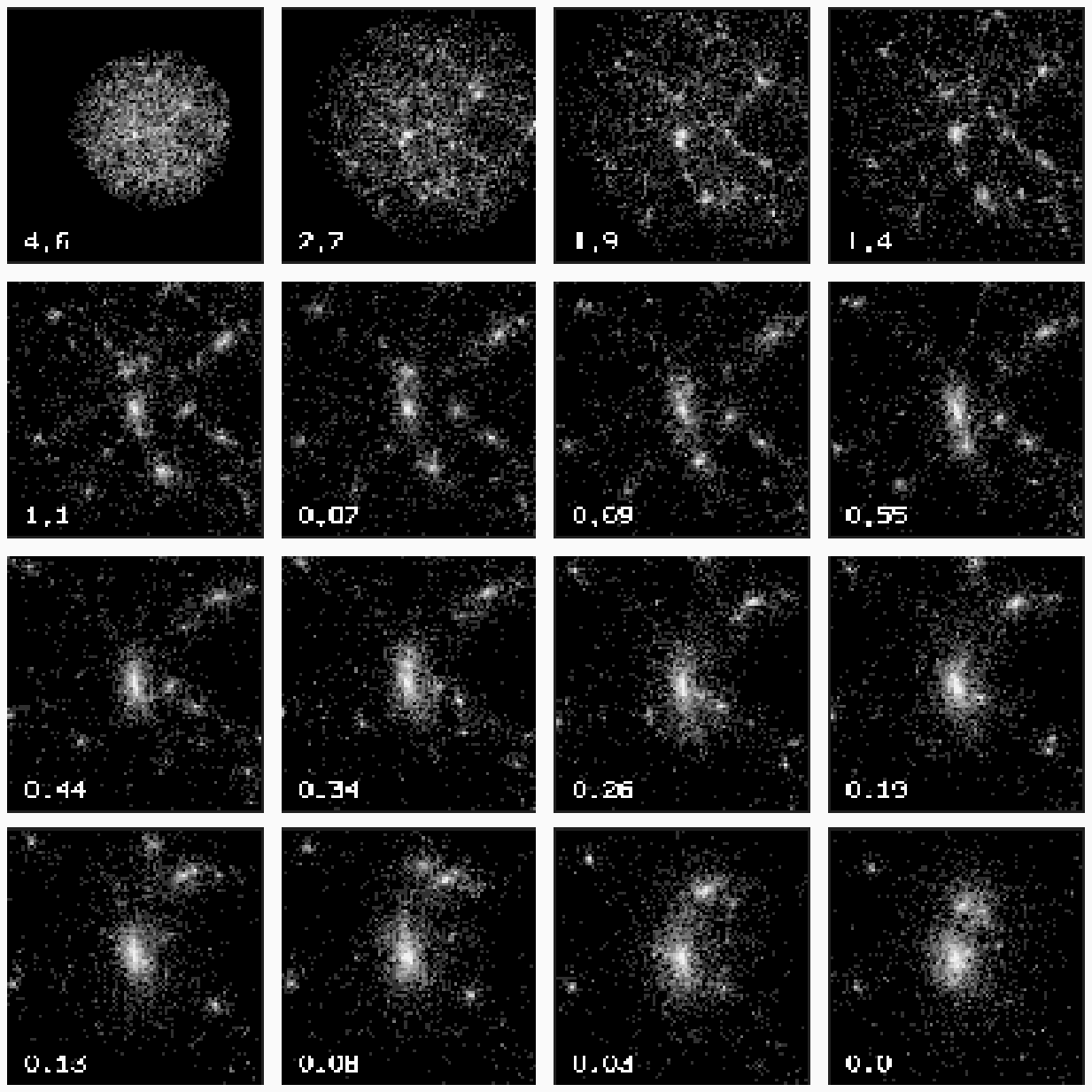}
\caption{Snapshots from Run 16M0  
at 16 different redshifts indicated by the number at 
lower-left of each panel. The length of the side for each panel 
is equal to $4r_{200}$ at $z_{\rm end}$.\label{fig1}}
}
\end{center}
\end{figure}

\begin{figure}
\begin{center}
{\leavevmode
\epsfxsize=15cm
\epsfbox{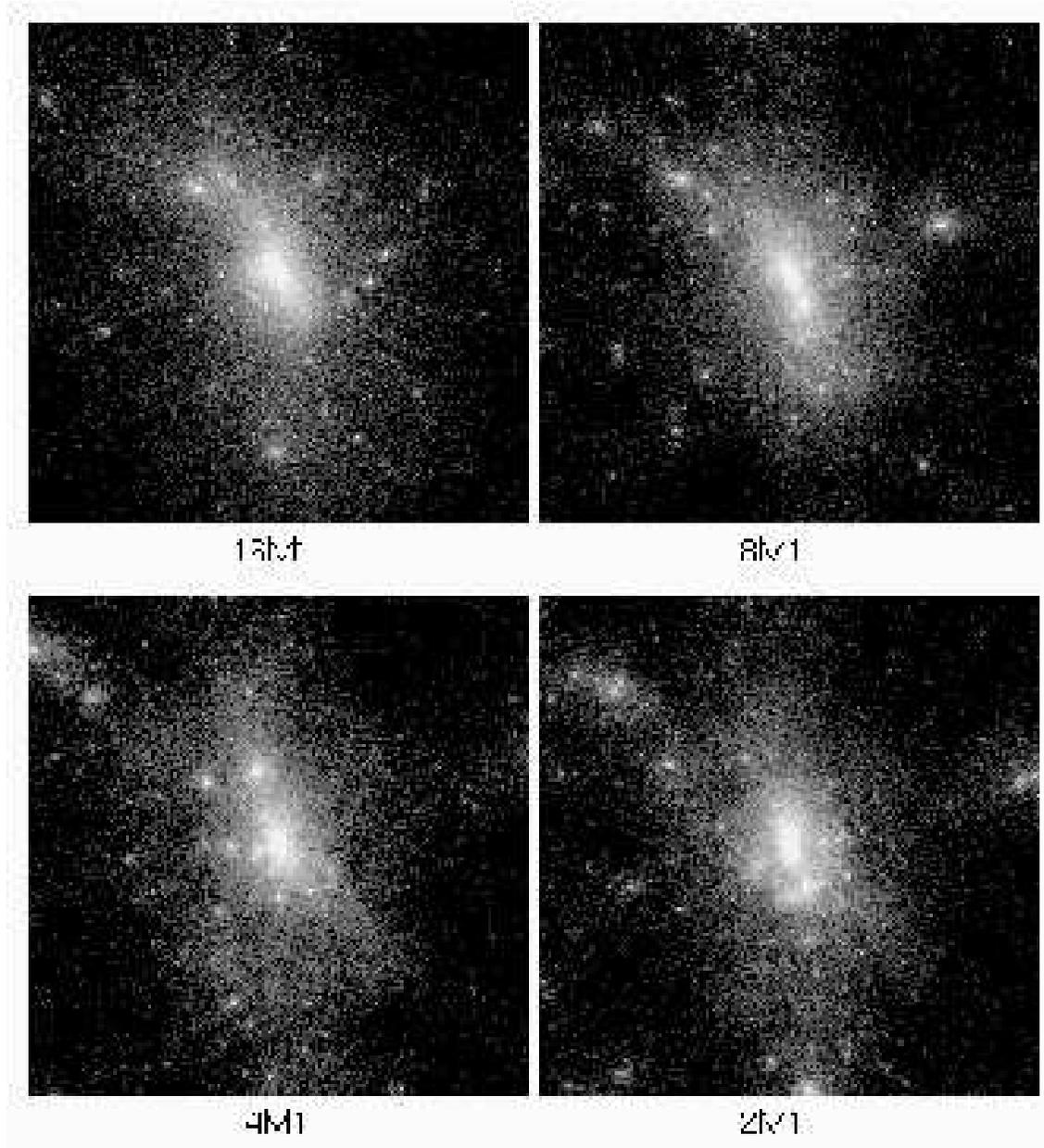}
\caption{Snapshots from Runs \{16,8,4,2\}M1 at the final redshifts. 
The length of the side for each panel is equal to $4r_{200}$ at $z_{\rm
end}$. \label{fig3}}
}
\end{center}
\end{figure}

Figure \ref{fig1} shows the particle distributions for Run 16M0 at 16
different redshifts.  For these plots, we shifted the origin of
coordinates to the position of the potential minimum so that the largest
halo is at the center of the panel.  Figure \ref{fig3} shows the
particle distribution for Runs \{16,8,4,2\}M1 together.  The phases of
waves for initial density field are the same for all these runs and only
the amplitude are different.  In Table 2, we summarized the radius
$r_{200}$, the mass $M_{200}$ and number of particles $N_{200}$ within
$r_{200}$, at $z_{\rm end}$. 
  
\begin{table}[h]
\caption{Halo properties at $z=z_{\rm end}$}
\begin{tabular}{lccr}
\hline
\hline
Run & $M_{200}$ ($M_{\odot}$) & $r_{200}$ (Mpc) & $N_{200}$ \\
\hline
16M0 & $2.6\times 10^{14}$ & 1.7 & 873170 \\
16M1 & $7.8\times 10^{14}$ & 2.4 & 1279383 \\
16M2 & $8.0\times 10^{14}$ & 2.4 & 1322351 \\
\hline
8M0 & $2.8\times 10^{13}$ & 0.48 & 745735 \\
8M1 & $9.0\times 10^{13}$ & 0.72 & 1186162 \\
8M2 & $7.7\times 10^{13}$ & 0.70 & 1015454 \\
\hline
4M0 & $2.7\times 10^{12}$ & 0.13 &  559563 \\
4M1 & $8.0\times 10^{12}$ & 0.20 & 846301\\
4M2 & $6.6\times 10^{12}$ & 0.22 & 697504 \\
\hline
2M0 & $6.6\times 10^{11}$ & 0.062 & 643151 \\
2M1 & $1.1\times 10^{12}$ & 0.085 & 957365 \\
2M2 & $1.0\times 10^{12}$ & 0.096 & 923545 \\
\hline
\end{tabular}
\end{table}

\subsection{Accuracy Criteria}
\label{secac}

In this study, we plot the density only for the radii unaffected by
numerical artifacts.  We used the following two criteria: (1) $t_{\rm
rel}(r)/t > 3$ and  (2) $t_{\rm dy}(r)/\Delta t > 40$. 
We obtained the criteria (1) and (2)
experimentally and the details of the experiments are discussed in
sections \ref{secactr} and \ref{secacdt}.  
We plot the
densities only if both criteria are satisfied.  Here, $t_{\rm
rel}(r)$ is the local two-body relaxation time defined by,
\begin{equation}
t_{\rm rel}={0.065v^3 \over G^2\rho m \ln(1/\varepsilon)},
\label{eqtr}
\end{equation}
(cf. Spitzer 1987) and $t_{\rm dy}(r)$ is the local dynamical time defined by
\begin{equation}
t_{\rm dy}=(G\bar{\rho})^{-1/2}
\label{eqdy}
\end{equation}
where $\bar{\rho}$ is the average density within radius $r$.  Using
these criteria we judge whether the density profile is unaffected by
numerical artifacts due to the two-body relaxation (1) and the step size
for the time integration (2).  

The inner limit of radii where the density is correctly calculated in
our simulations is $(0.01-0.02) r_{200}$ at the final redshifts.  In
most cases, the criterion (1) for the two-body relaxation determines the
limit radius for reliability. 

\subsubsection{Criterion for two-body relaxation}
\label{secactr}

\begin{figure}
\begin{center}
{\leavevmode
\epsfxsize=10cm
\epsfbox{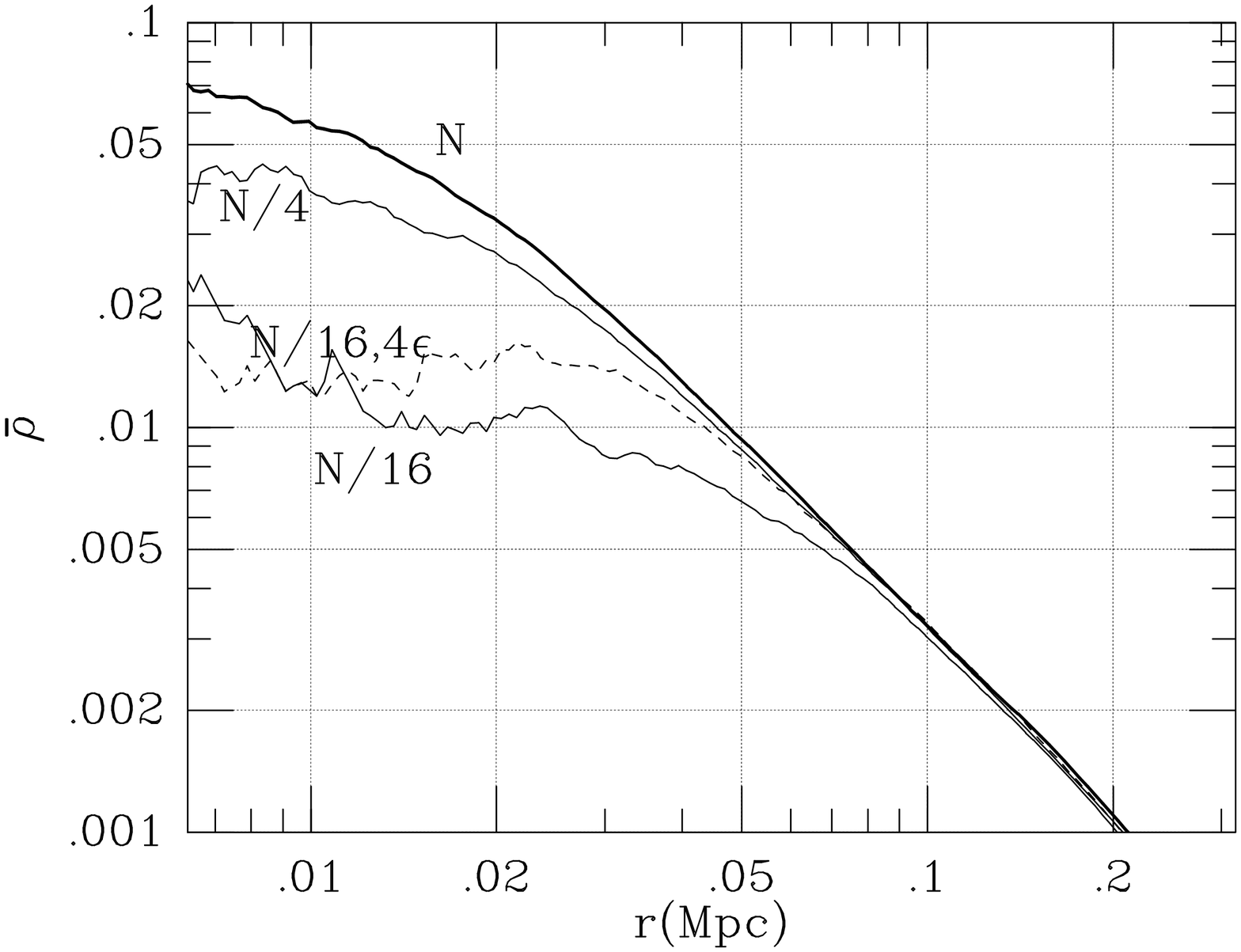}
\caption{Averaged density profiles for Run 16M0 and the same model as
Run 16M0 but with $N$/4, $N$/16, and $N$/16 and 4$\varepsilon$, where
$N$ and $\varepsilon$ are the number of particles and softening
parameter for Run 16M0.  The model with 4$\varepsilon$ is plotted in the
dashed line.  The unit of density is $M_{\odot}$/pc$^3$. 
\label{figac1}}}
\end{center}
\begin{center}
{\leavevmode
\epsfxsize=10cm
\epsfbox{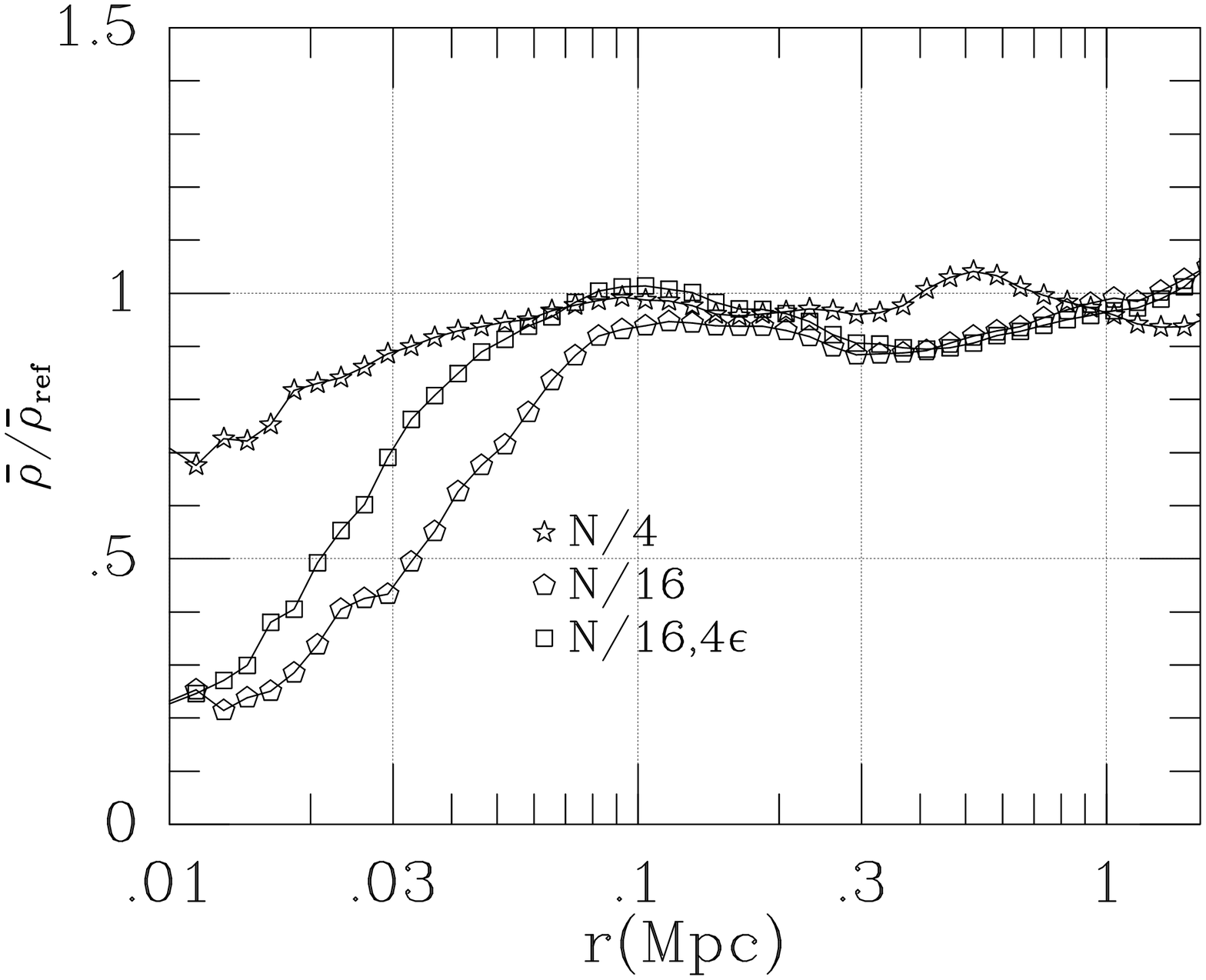}
\caption{The ratio of the averaged density for the same models as Fig
\ref{figac1} to that for the reference run (Run 16M0). 
\label{figac2}}}
\end{center}
\end{figure}

In this subsection, we evaluate the criterion, $t_{\rm rel}(r)/t > 3$,
to distinguish the numerical artifact due to the two-body relaxation. 
In order to see the effect of two-body relaxation, we calculated the
same model as Run 16M0 but with several different values for total
number of particles ($N$) and the softening size ($\varepsilon$).  In
Figure \ref{figac1}, we plot the final average density profiles ${\bar
\rho}$ for three simulations with $N/4$, $N/16$, and $N/16$ and
$4\varepsilon$, where $N$ and $\varepsilon$ mean values used in Run
16M0.  Otherwise stated, we used the same simulation parameter as in Run
16M0.  We can see that the central density depends on $N$ rather
strongly, and is lower for lower number of particles.  For the same
value of $N$, larger softening has small but clear effect of increasing
the central density. 

Figure \ref{figac2} shows the ratio, ${\bar \rho}/{\bar \rho_{\rm
ref}}$, where ${\bar \rho_{\rm ref}}$ is the averaged density of the
reference run in which the effect of the two-body relaxation is
smallest.  Here, we used Run 16M0 as the reference run.  In Figure
\ref{figac3}, we show the ratio ${\bar \rho}/{\bar \rho_{\rm ref}}$
plotted as a function of the ratio of the local two-body relaxation time
$t_{\rm rel}(r)$ defined by (\ref{eqtr}) to simulation period $t$.  Note
that $t_{\rm rel}(r)$ is monotonous and increasing function of $r$. 
Therefore, smaller $t_{\rm rel}(r)$ means smaller $r$.  We can see that
the density tends to go below the reference value if $t<3 t_{\rm
rel}(r)$.  The difference between the reference run and runs with
smaller number of particles is insignificant if $ t<3 t_{\rm rel}(r)$. 
From this result, we adapt $t_{\rm rel}(r)/t > 3$ as the criterion for
the two-body relaxation. 

\subsubsection{Criterion for time integration}
\label{secacdt}

\begin{figure}
\begin{center}
{\leavevmode
\epsfxsize=10cm
\epsfbox{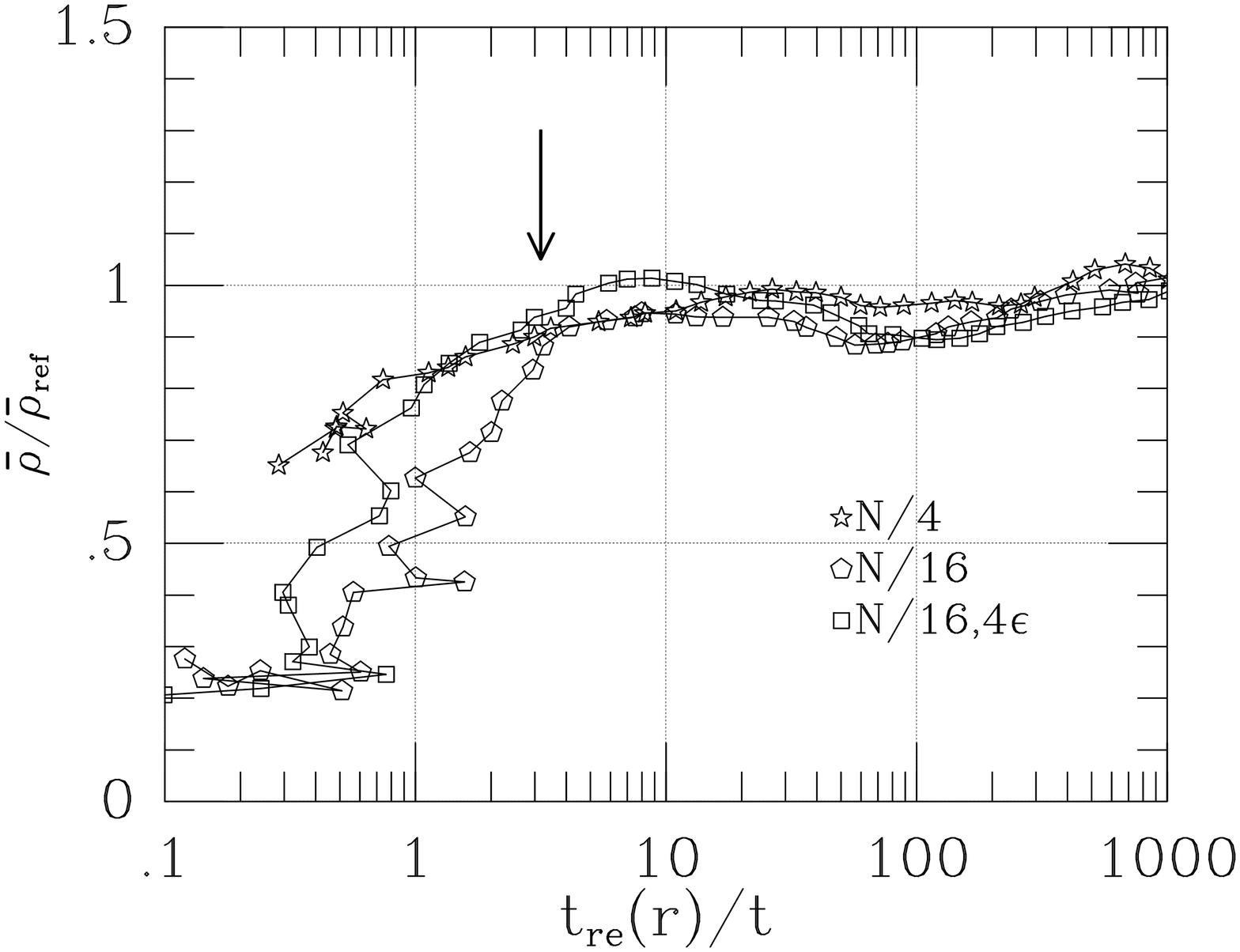}
\caption{Same as Figure \ref{figac2}, but as a function of 
the ratio of the local relaxation time to the simulation period, 
$t_{\rm re}(r)/t$. The arrow indicates the accuracy criterion 3. 
\label{figac3}}}
\end{center}
\begin{center}
{\leavevmode
\epsfxsize=10cm
\epsfbox{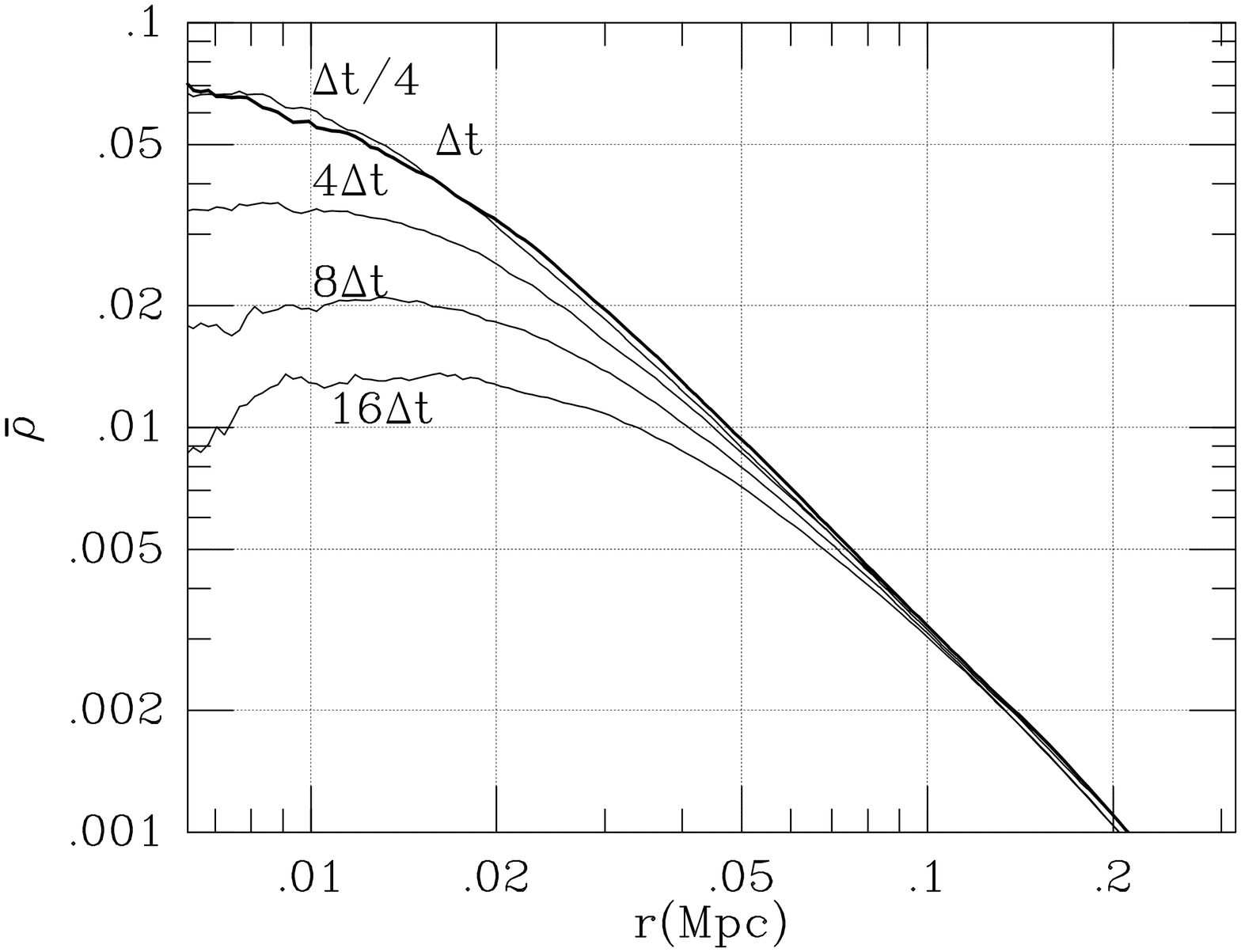}
\caption{Averaged density profile for Run 16M0 in the thick line and the
same model as Run 16M0 but with 4, 8, and 16 times larger and 4 times
smaller time stepsize in the thin line.  The unit of density is
$M_{\odot}$/pc$^3$. 
\label{figac4}}}
\end{center}
\end{figure}

\begin{figure}
\begin{center}
{\leavevmode
\epsfxsize=10cm
\epsfbox{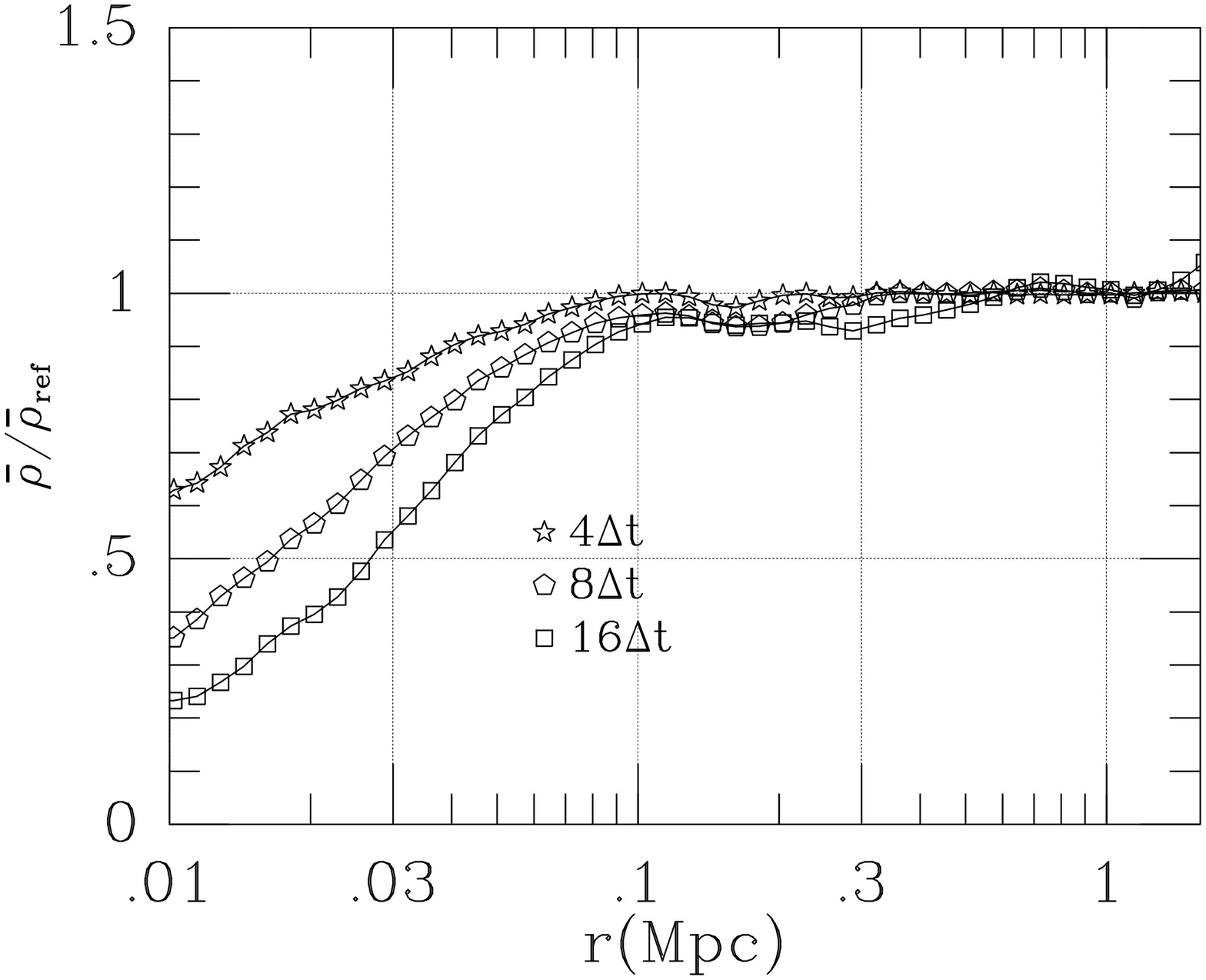}
\caption{The ratio of the averaged density 
for the same model as Fig \ref{figac4} for the reference run (Run 16M0). 
\label{figac5}}}
\end{center}
\begin{center}
{\leavevmode
\epsfxsize=10cm
\epsfbox{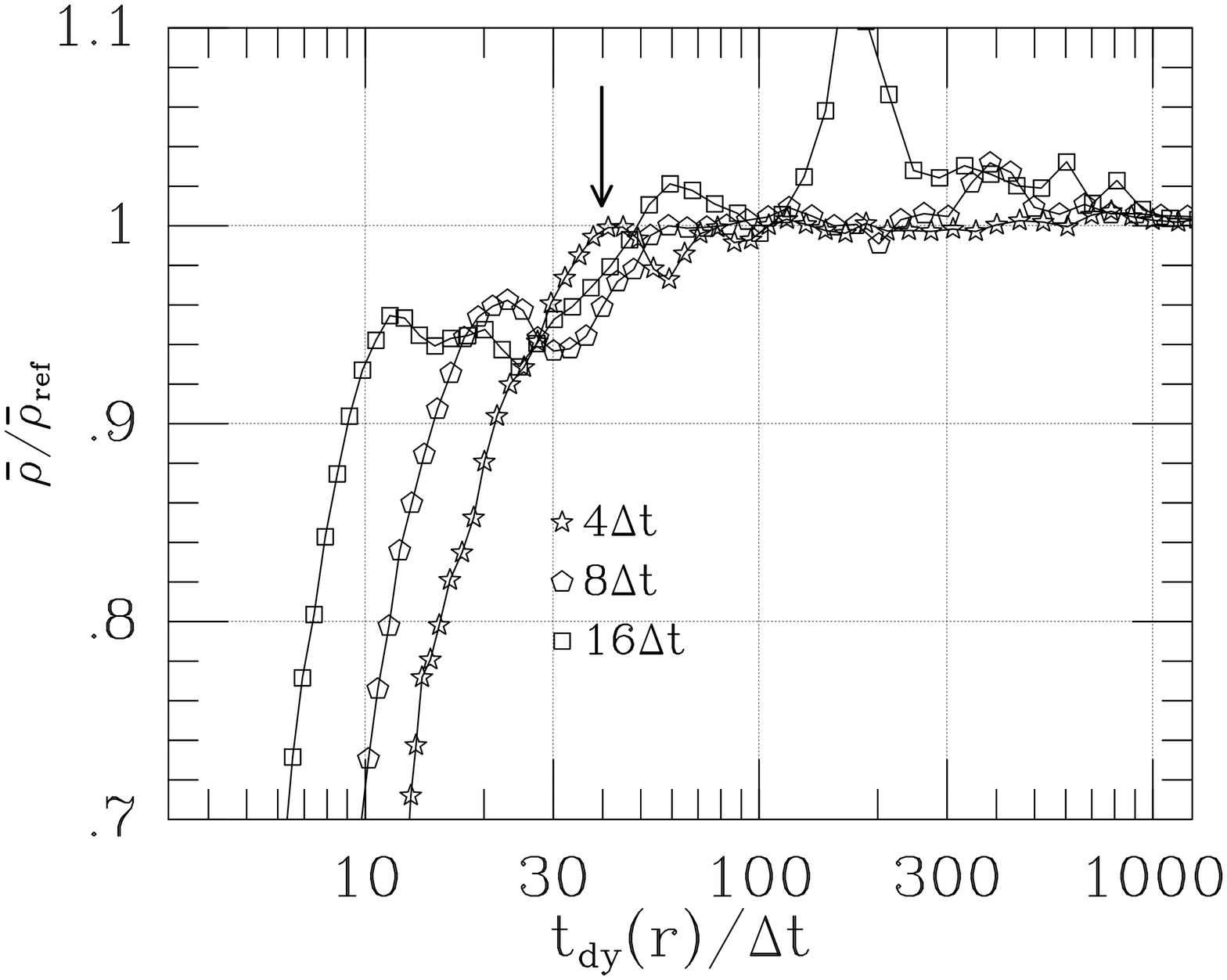}
\caption{Same as Figure \ref{figac5}, but as a function of the ratio of
the local dynamical time to time step size, $t_{\rm dy}(r)/\Delta t$. 
The arrow indicates the accuracy criterion 40. 
\label{figac6}}}
\end{center}
\end{figure}

In this subsection, we evaluate the criterion, $t_{\rm dy}(r)/\Delta t
> 40$, to distinguish the numerical artifact due to large step size of
time integration.  In order to see the effect of large step size, we
calculated the same model as Run 16M0 but with larger time step size
($\Delta t$).  In Figure \ref{figac4}, we plot the final average density
profiles ${\bar \rho}$ for three simulations with $4\Delta t$, $8\Delta
t$, and $16\Delta t$, where $\Delta t$ means values used in Run 16M0. 
Otherwise stated, we used the same simulation parameter as in Run 16M0. 
We can see that the central density depends on $\Delta t$ rather
strongly and is lower for larger $\Delta t$.  Figure \ref{figac5} shows
the ratio, ${\bar \rho}/{\bar \rho_{\rm ref}}$.  Here, we used Run 16M0
as the reference run. 

In Figure \ref{figac6}, we show the ratio ${\bar \rho}/{\bar \rho_{\rm
ref}}$ plotted as a function of ratio of the local dynamical time
$t_{\rm dy}$ defined by (\ref{eqdy}) to the time step size $\Delta t$. 
Note that $t_{\rm dy}(r)$ is monotonous and increasing function of $r$. 
Therefore, smaller $t_{\rm dy}(r)$ means smaller $r$.  We can see that
the density tends to go below the reference value if $t_{\rm dy}(r)< 40
\Delta t$.  From this result, we adapt $t_{\rm dy}(r)/\Delta t > 40 $
as the criterion for time integration. 

Note that the number 40 is applicable only to the integration scheme we
used: the leapfrog scheme integrated in physical space with constant
time step size.  This scheme has good characteristics such as the
time-reversibility and symplecticity.  The number 40 should increase
when variable stepsize is used or the system is integrated in comoving
space (where acceleration depends on velocity). 

As a result of adapting this criterion, the total number of time steps
to integrate in Hubble time becomes about 8000 in our simulation.  The
number is a little smaller than that reported in previous simulation
(ex.  $\sim 50000$, Moore et al.  1999).  We also calculated the same
model as Run 16M0 but with 4 times many time steps.  We confirmed that
the density profile, shape, and anisotropy parameter do not change, and
the number of timestep 8000 is enough.  In figure \ref{figac4} we show
the density profile. 

\subsubsection{Other numerical effects}

\begin{figure}
\begin{center}
{\leavevmode
\epsfxsize=12cm
\epsfbox{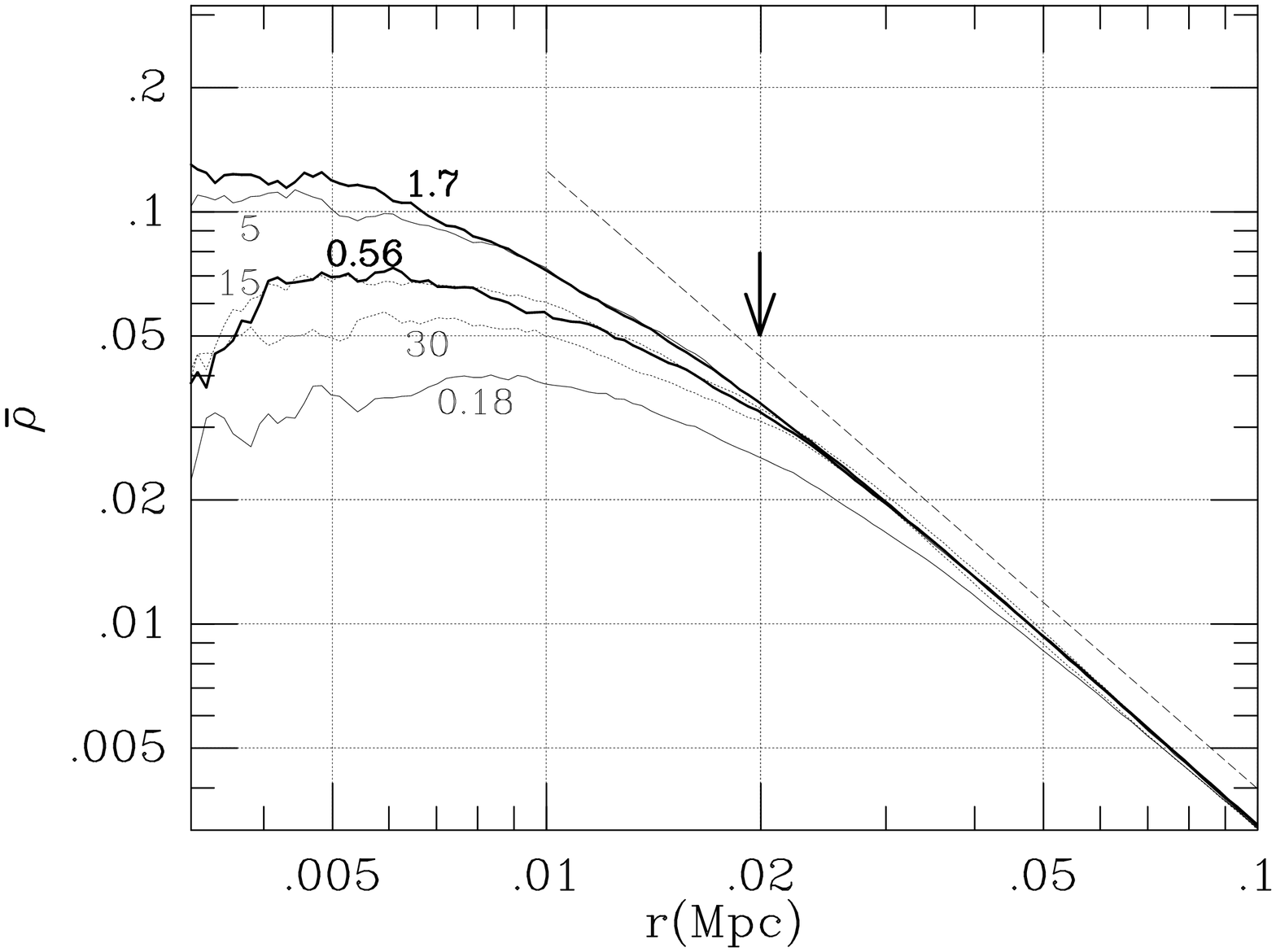}
\caption{
Averaged density profiles for Run 16M0 ($\varepsilon=0.56$kpc) and the
same model as Run 16M0 with $\varepsilon=$0.18, 1.7, 5, 15, 30 kpc,
where $\varepsilon$ is the softening length.  The unit of density is
$M_{\odot}$/pc$^3$.  The numbers beside the lines indicates the
softening length.  The profiles for Run 16M0 and the models with
$\varepsilon=1.7$kpc are plotted in the thick lines.  The dashed lines
indicates the density profile proportional to $r^{-1.5}$.  The arrow
indicates the critical radius defined by the accuracy criteria (1) and
(2). 
\label{figeps}}
}
\end{center}
\end{figure}

The potential softening also affects the density profile.  In order to
see the effect of the potential softening, we calculated the same model
as Run 16M0 but with different softening length ($\varepsilon$).  In
Figure \ref{figeps}, we plot the final average density profiles ${\bar
\rho}$ for six models with $\varepsilon$ = 0.18, 0.56, 1.7, 5, 15, 30
kpc.  Except for the softening length, we used the same simulation
parameter as in Run 16M0.  In Figure \ref{figeps} we can see that the
central density is lower for both smaller and larger $\varepsilon$.  The
former is because the two-body relaxation effect is stronger and the
time integration is less accurate for smaller $\varepsilon$.  The latter
is because the potential softening itself affects the density structure
for larger $\varepsilon$.  The potential softening, therefore, should be
set in the intermediate range. 

The potential softening for Run 16M0 ($\varepsilon=$0.56kpc) is in the
intermediate range, though it is not optimal.  It does not affect the
density profile outside of 20 kpc, which is the critical radius defined
by the accuracy criterion (1) and (2).  The ratio of the softening
length to the critical radius in other runs are similar to that in Run
16M0. 

We made sure that the accuracy of the BH tree code did not influence the
structure in the range where the above criteria are satisfied, by
re-simulating the same initial model as used in FM97, in which the
direct summation is used.  We found no systematic difference in the
results.  Therefore, the accuracy of BH tree-code is okay. 

\subsection{Density Profiles}

\begin{figure}
\begin{center}
{\leavevmode
\epsfxsize=13cm
\epsfbox{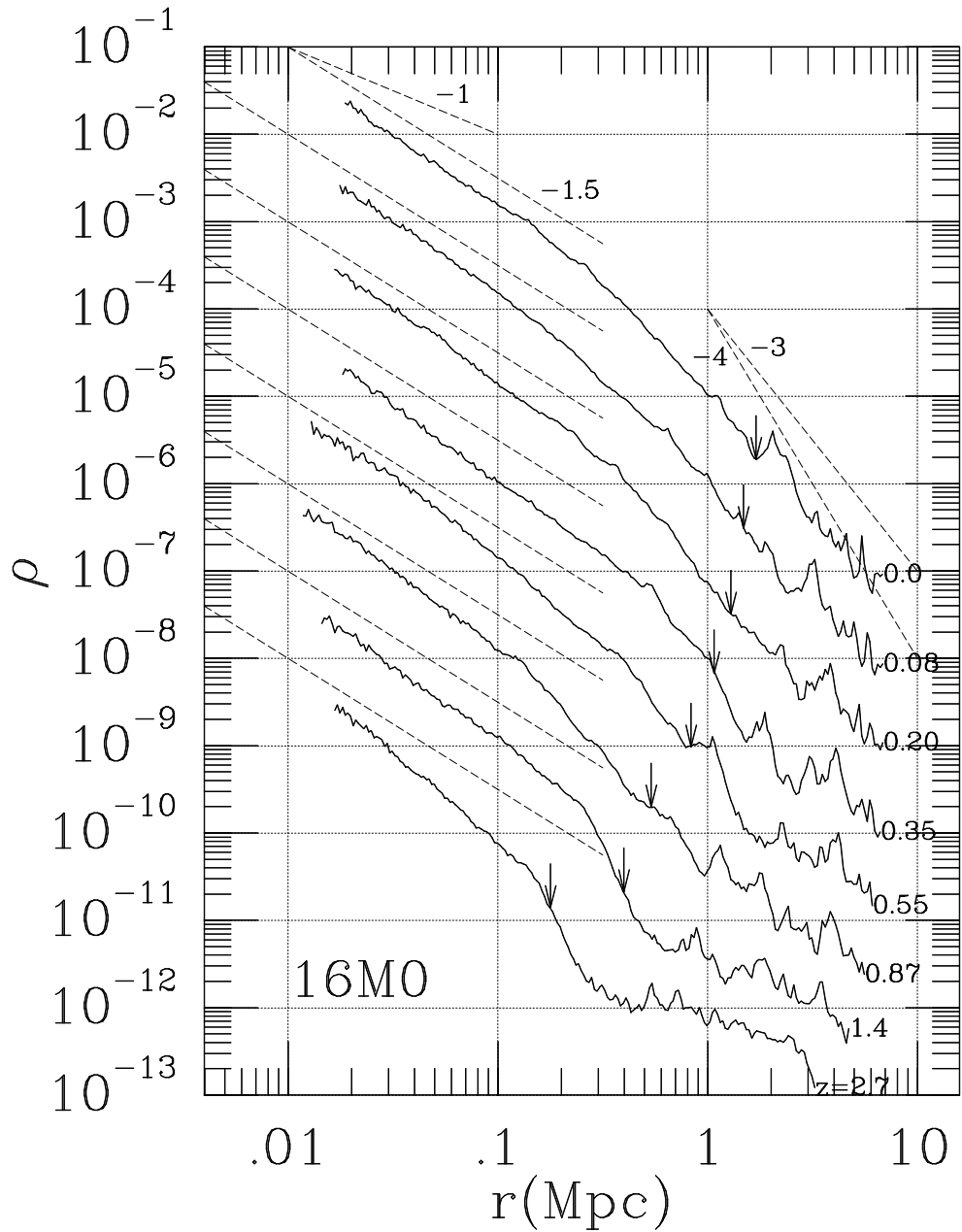}
\caption{Evolution of density profile of the halo for Run 16M0.  
The unit of density is $M_{\odot}$/pc$^3$. 
The profiles are vertically shifted downward by $7, 6,...,1,0$ dex from
the bottom to the top.  The arrows indicates $r_{200}$.  The numbers
near the dashed lines indicate the power index of those lines.  The numbers on
the left of the profiles indicate the redshift.  \label{fig4}}
}
\end{center}
\end{figure}

\begin{figure}
\begin{center}
{\leavevmode
\epsfxsize=13cm
\epsfbox{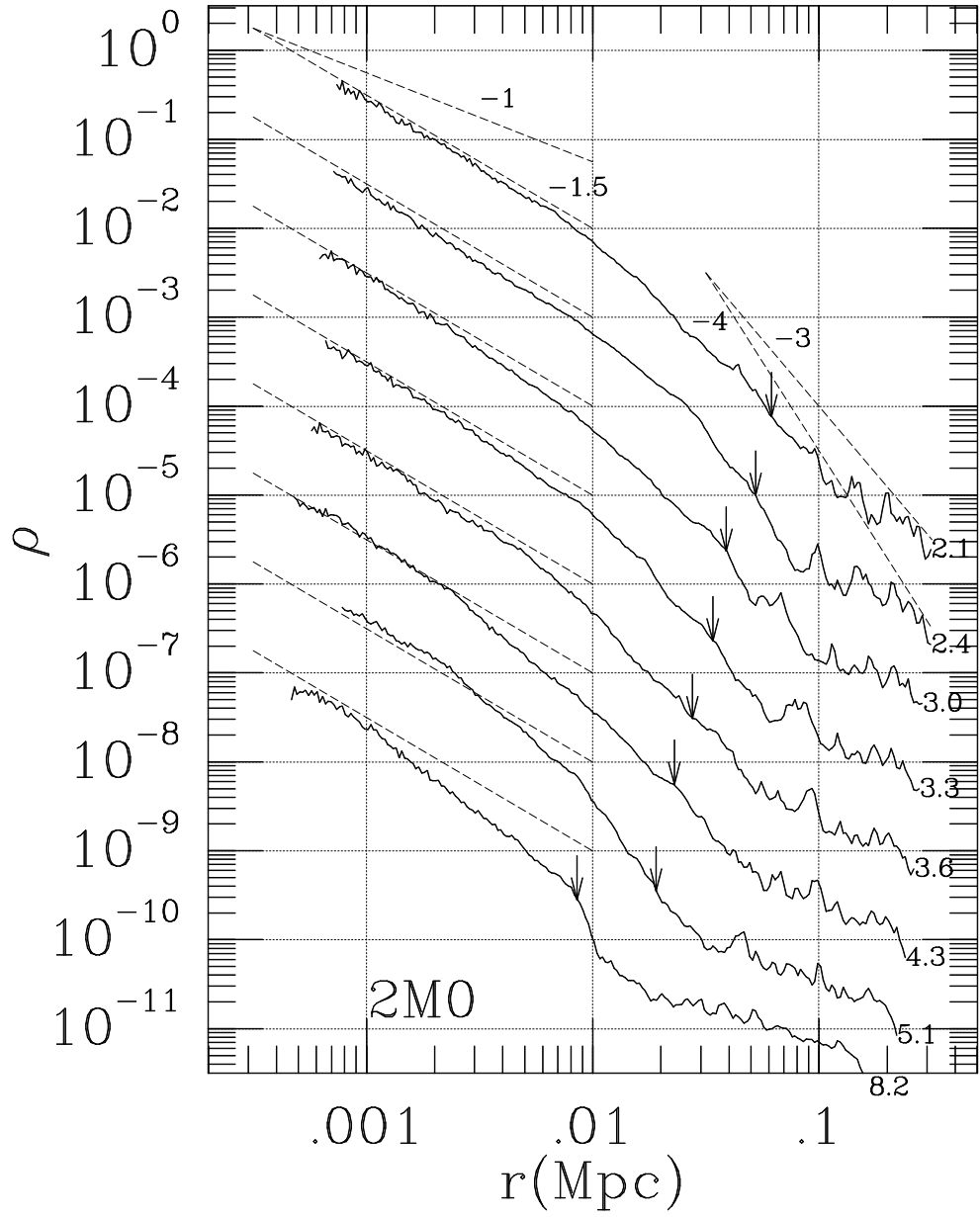}
\caption{Same as Figure \ref{fig4}, but for Run 2M0.\label{fig5}}
}
\end{center}
\end{figure}

\begin{figure}
\begin{center}
{\leavevmode
\epsfxsize=17cm
\epsfbox{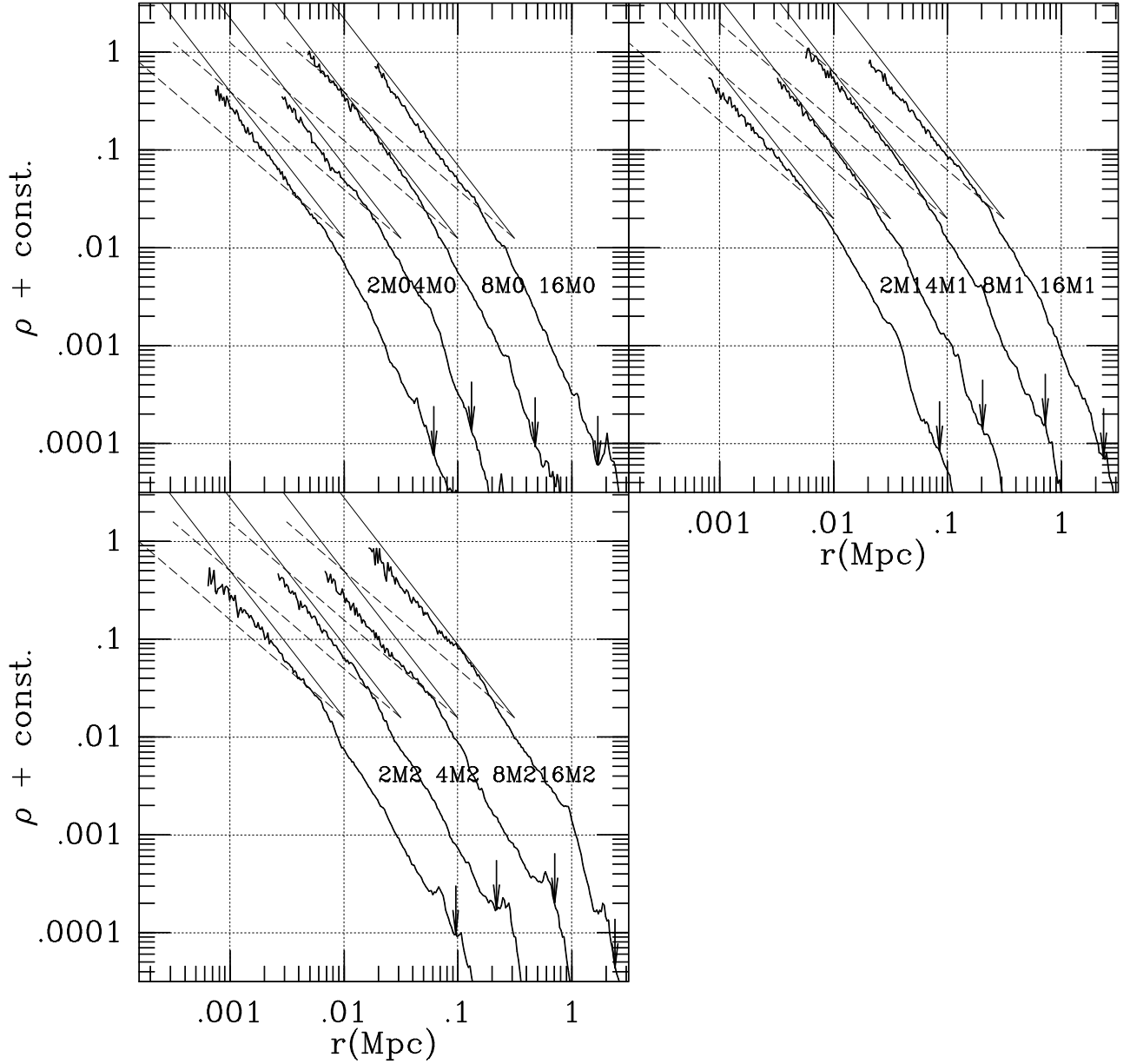}
\caption{Density profile of the halo for all
Runs at $z_{\rm end}$.  The unit of density is $M_{\odot}$/pc$^3$.  The
profile for Run 16M\{0,1,2\}, 8M\{0,1,2\}, and 4M\{0,1,2\} are vertically
shifted upward by $1.5$, $1$, $0.5$ dex, respectively.  The arrows
indicate $r_{200}$.  The thin solid and dashed lines indicate the
densities proportional to $r^{-1.5}$ and $r^{-1}$.  \label{fig6}}
}
\end{center}
\end{figure}

Figure \ref{fig4} shows the evolution of density profiles for Run 16M0. 
The position of the center of the halo was determined using the
potential minimum and the density is averaged over each spherical shell
whose width is $\log_{10}(\Delta r)=0.0125$.  Figure \ref{fig5} is for
Run 2M0.  For the illustrative purpose, the densities are shifted
vertically.  Figure \ref{fig6} show the density profiles at $z_{\rm
end}$ for all Runs. 

In all runs we can see the central density cusps approximately
proportional to $r^{-1.5}$.  In other words, the power of the cusp is
$-1.5$ and is independent of halo mass, which is consistent with the
result of Moore et al.(1999).  In the outer region, the density profiles
are very similar for all runs.  The dependence of the power-law index of
the inner cusp on the halo mass observed by Jing and Suto (2000) was not
reproduced in our simulations.  Even if we take into account the
run-to-run variation, the dependence on mass in our results is in the
opposite direction compared to that of Jing and Suto (2000).  In the
following subsections, we discuss the self-similar growth of the halo
(section \ref{sec_self}), the universality of the profile (section
\ref{sec_univ}), and the mechanism for the self-similar growth (section
\ref{sec_mec}). 

\subsection{Self-Similar Evolution}
\label{sec_self}

Figure \ref{fig7} shows the growth of the halo, without the vertical
shift. In this figure it is clear that the halo grows in a
self-similar way, keeping the density of the central cusp region
constant. 

If the evolution is self-similar, we 
can write the density as 
\begin{eqnarray}
\rho(r,t) & = & \rho_{\dagger}(M)\rho_{\ast}(r_{\ast})\label{eq1} \\
r_{\ast}& = & r/r_{\dagger}(M) 
\end{eqnarray}
Here, we write $\rho_{\dagger}(M)$ and $r_{\dagger}(M)$ as a function of
the mass of the halo $M$, instead of the time.  The self-similar profile
itself should have the central cusp of $\rho_{\ast}(r_{\ast})\propto
r_{\ast}^{n}$.  The actual profile at the cusp region satisfies
$\rho(r)=Cr^{n}$, with $C$ constant in time.  Therefore,
$\rho_{\dagger}$ and $r_{\dagger}$ should satisfy $\rho_{\dagger}\propto
r_{\dagger}^n$.  If we write $\rho_{\dagger}$ and $r_{\dagger}$ as
function of $M$, we have
\begin{eqnarray}
\rho_{\dagger}(M) & = & \rho_{00}\displaystyle\left({M \over M_{00}}\right)^{n\over 3+n} \\
r_{\dagger}(M) & = & r_{00}\left({M \over M_{00}}\right)^{1\over 3+n} 
\label{eq2}
\end{eqnarray}
where $r_{00}$, $\rho_{00}$, $M_{00}$ are constants and $n$ is the power-law
index of the cusp given by $\rho \propto r^{n}$.  This self-similarity is illustrated 
in Figure \ref{figself}. If we set $n=-1.5$ from the simulations, 
we obtain 
\begin{eqnarray}
\rho_{\dagger}(M) & \propto & M^{-1}\\
r_{\dagger}(M) & \propto & M^{2\over 3}. 
\end{eqnarray}

In Figure \ref{fig8}, we plot $\rho_{\ast}$ defined through equations
(\ref{eq1})-(\ref{eq2}) as a function of $r_{\ast}$ Here, we took
$M_{00}=10^{14}M_{\odot}$, $r_{00}=0.2$ Mpc, and $\rho_{00}=7\times
10^{-4} M_{\odot}/$pc$^3$.  We plot four density profiles at different
values of the redshift $z$.  We set $n=-1.5$ for all runs and use
$M_{200}$ as the total mass.  In this figure, we can see that the
density profiles of the same halo at different times show very good
agreement to each other, which means the density structure evolves
self-similarly, though in outer region a degree of overlapping becomes
worse.  Therefore, Figure \ref{fig8} demonstrates that our assumption of
the self-similarity is justified. 

\subsection{Universality}
\label{sec_univ}

In this subsection we discuss the universality of the density profile. 
Using a non-dimensional free parameter $\delta$, we define new
non-dimensional variables expressed as 
\begin{eqnarray} 
\rho_{\ast\ast} & = & \rho_{\ast}\delta^{-1} \\ 
r_{\ast\ast} & = & r_{\ast}\delta^{1\over 3} 
\end{eqnarray} 
Figure \ref{fig9} shows $\rho_{\ast\ast}$ $r_{\ast\ast}$ of all Runs
 at $z=z_{\rm end}$.  The values of
$\delta^{-1}$ are 1.0, 0.4, and 0.6 for Run 16M\{0,1,2\}, 2.5, 1.0 and
3.0 for Run 8M\{0,1,2\}, 10.0, 3.0 and 6.0 for Run 4M\{0,1,2\}, 
and 35.0, 12.0 and 30.0 for Run 2M\{0,1,2\}. We can
see that the 12 density structures agree very nicely, which means they
are universal.  In principle, any scaling on the $r$-$\rho$ plane can be
expressed using two parameters.  We used the total mass $M$ as one of
two parameters to express the self-similarity discussed in section
\ref{sec_self}.  The parameter $\delta$ corresponds to the other
freedom.  The value of $\delta$ is considered to reflect an amplitude of
the density fluctuation at the collapse. 

We attempted to fit the density structure to several profiles proposed
in earlier studies.  Figure \ref{fig10} and \ref{fig11} show the profile
proposed by Moore et al.  (1999) and by NFW. 
The function forms are given by
$\rho_{\ast\ast}=r_{\ast\ast}^{-1.5}(1+r_{\ast\ast}^{1.5})^{-1}$ and by
$\rho_{\ast\ast}=10(0.5r_{\ast\ast})^{-1}(1+0.5r_{\ast\ast})^{-2}$,
respectively.  Our simulation results agree with the profile proposed by
Moore et al.(1999) very well, while the agreement with the NFW profile
is not so good. 

In summary, the density structure of simulated halos is well expressed by
\begin{equation} 
{\rho(r)\over \rho_0} (= \rho_{\ast\ast})
= \displaystyle{1 \over (r/r_0)^{1.5}[1+(r/r_0)^{1.5}]} 
\left(= \displaystyle{1 \over r_{\ast\ast}^{1.5}[1+r_{\ast\ast}^{1.5}]} \right)
\end{equation} 
where 
\begin{eqnarray}
\rho_0 &=& 7\times 10^{-4} \cdot \delta 
\displaystyle\left({M\over 10^{14}M_{\odot}}\right)^{-1}
\quad (M_{\odot}{\rm /pc^3})\\
r_0 &=& 0.2 \cdot \delta^{-{1 \over 3}} 
 \displaystyle\left({M\over 10^{14}M_{\odot}}\right)^{2 \over 3} 
\quad ({\rm Mpc}) 
\end{eqnarray} 
where again $M$ is the total mass of halos, $\delta$ is a
non-dimensional free parameter.  The free parameter $\delta$ is constant
during evolution of a halo. 

\subsection{Formation Process of The Central Cusp}
\label{sec_mec}

In this subsection, we show that the central cusp grows through the
accretion of the disrupted smaller halos by a larger halo.  In the
bottom-up structure formation a typical halo grows through repeated
merging of smaller halos.  In the CDM hierarchical clustering, the
larger halo typically has a denser central region than the smaller halo
has.  Therefore, the smaller halo is disrupted by the tidal field of the
larger halo and the matter from the smaller halo is scattered around,
when two halos merge.  On the other hand, the central region of the
larger halo survives the merging process more or less intact. 

\begin{figure}
\begin{center}
{\leavevmode
\epsfxsize=12cm
\epsfbox{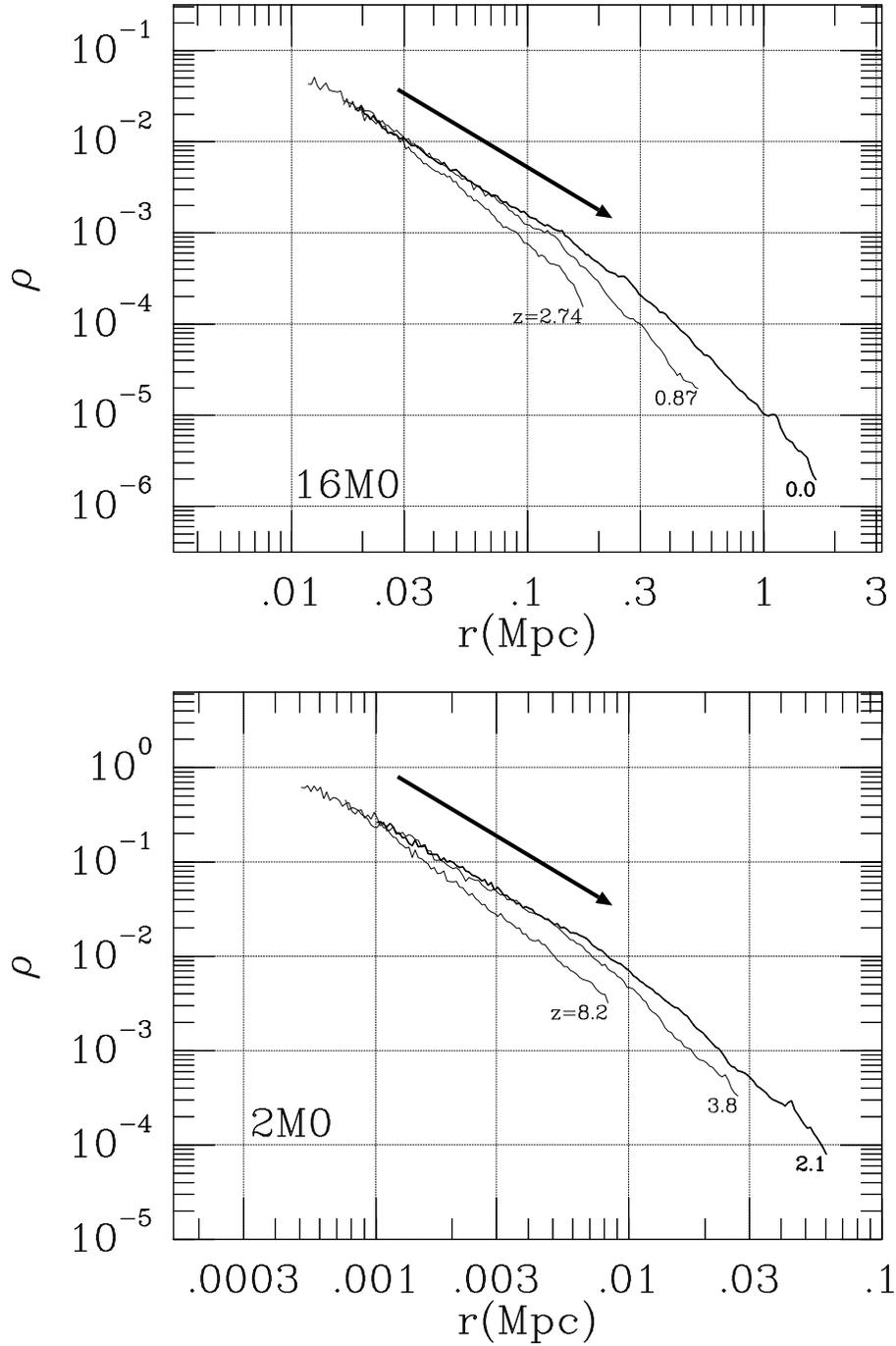}
\caption{
Evolution of the density profile of the halo for 
Runs 16M0 and 2M0 with no vertical shift.  The unit of density is
$M_{\odot}$/pc$^3$.  \label{fig7} }
}
\end{center}
\end{figure}

\begin{figure}
\begin{center}
{\leavevmode
\epsfxsize=12cm
\epsfbox{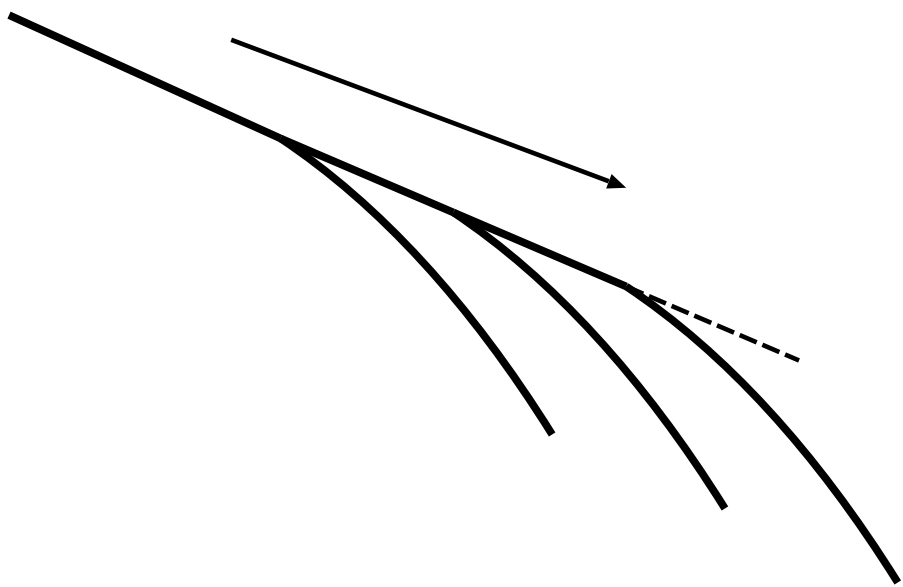}
\caption{
Self-Similar Evolution.\label{figself} }
}
\end{center}
\end{figure}

\begin{figure}
\begin{center}
{\leavevmode
\epsfxsize=12cm
\epsfbox{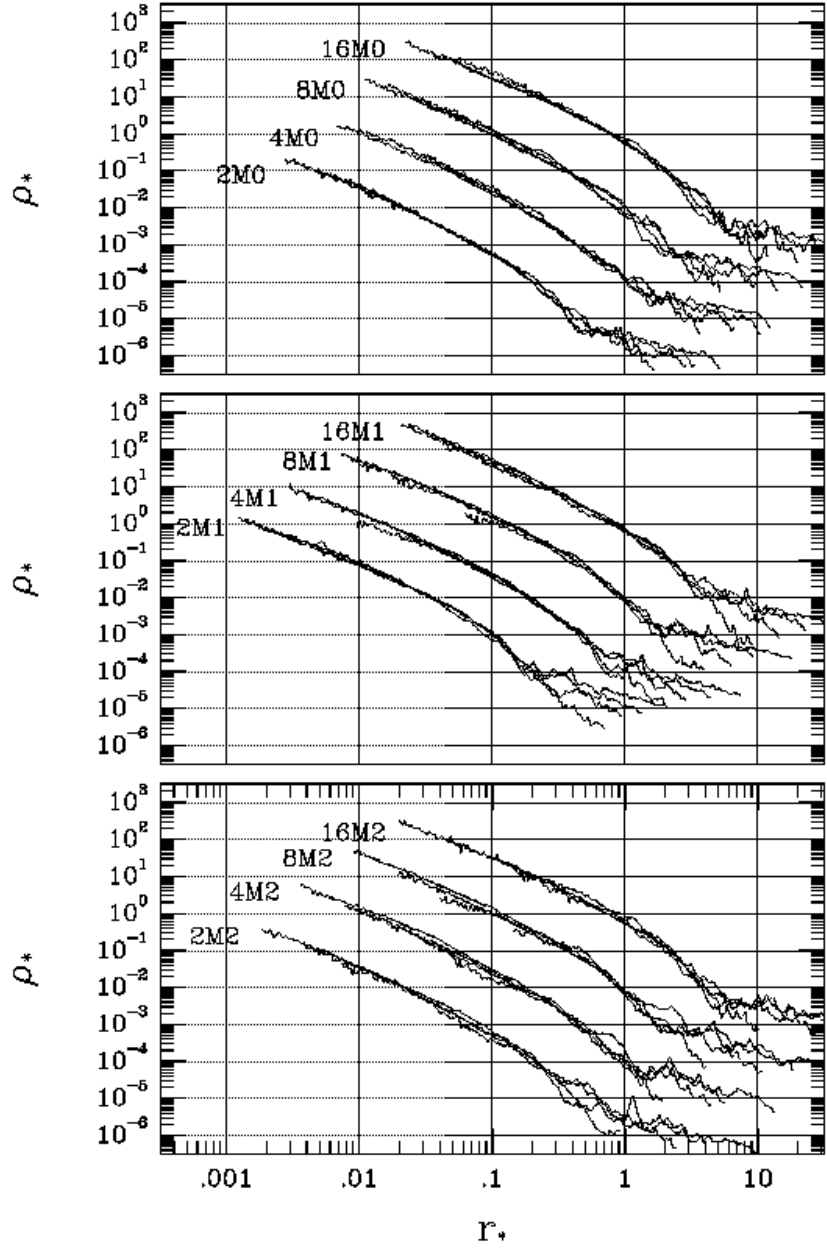}
\caption{Self-similar evolution of the density profile. 
The scaled densities $\rho_{\ast}$ are plotted 
as a function of the scaled radius $r_{\ast}$. 
The profiles for Run 8M\{0,1,2\}, 4M\{0,1,2\} and 2M\{0,1,2\} 
are vertically shifted downward by 0.5, 1, 1.5 dex and
horizontally shifted to the left by 0.5, 1, 1.5 dex, respectively.
\label{fig8}}
}
\end{center}
\end{figure}

\begin{figure}
\begin{center}
{\leavevmode
\epsfxsize=15cm
\epsfbox{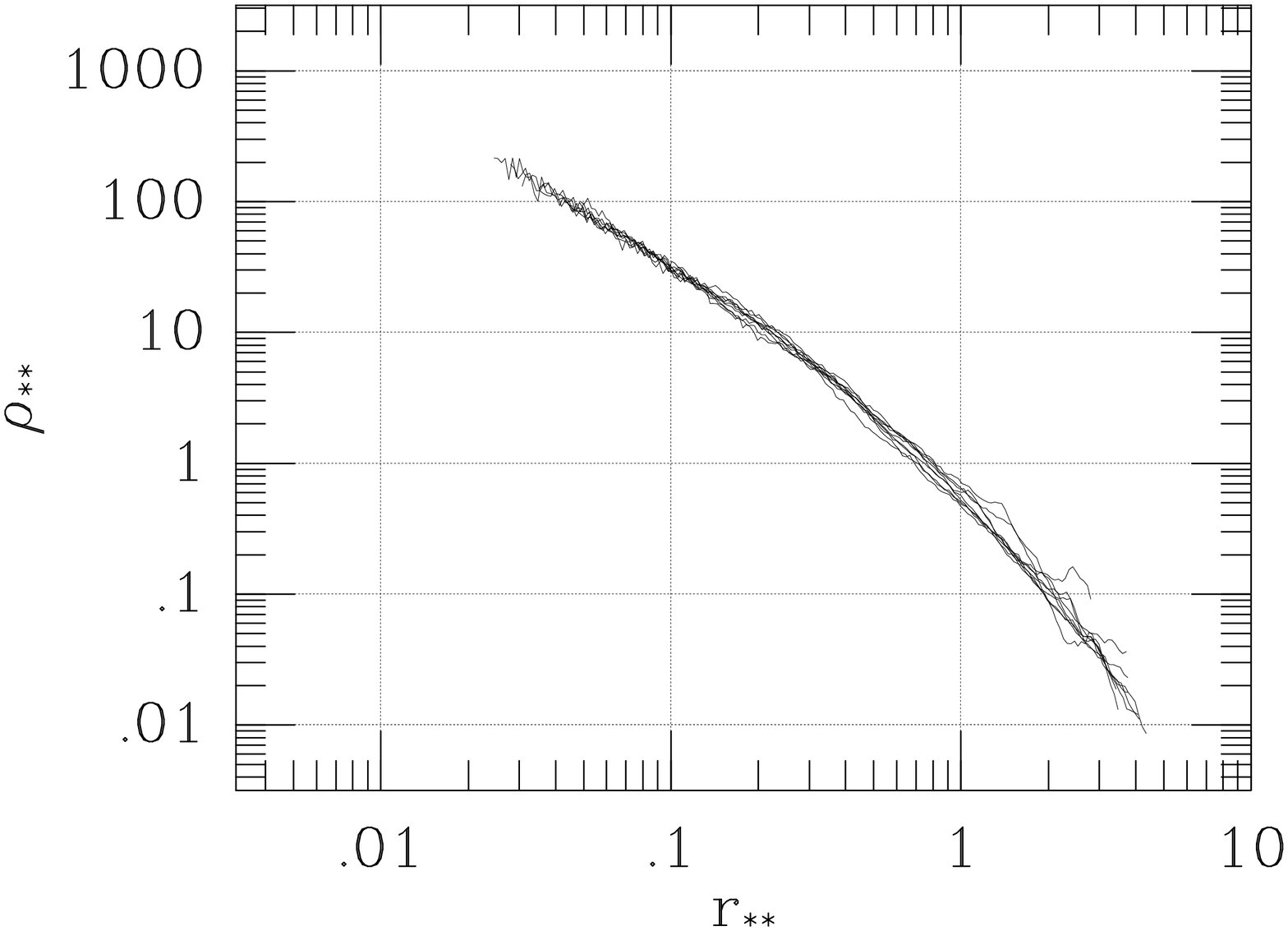}
\caption{
Universality in the density structure.  The scaled densities
$\rho_{\ast\ast}$ are plotted as a function of the scaled radius
$r_{\ast\ast}$. 
\label{fig9}}}
\end{center}
\end{figure}

\begin{figure}
\begin{center}
{\leavevmode
\epsfxsize=13cm
\epsfbox{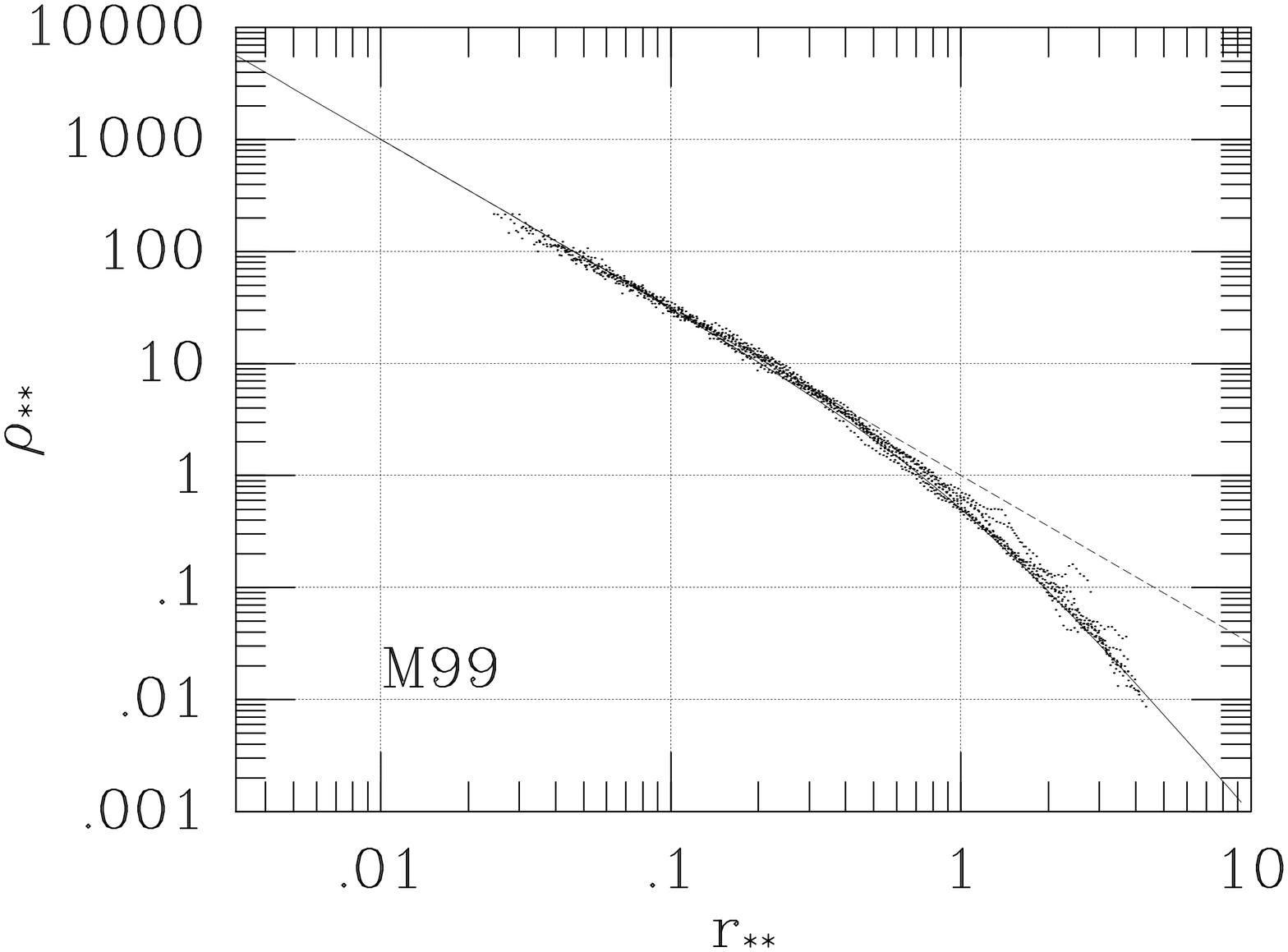}
\caption{
Fitting of the density structure by the profile proposed by Moore et al. (1999) (solid curve). 
The dashed line indicates $\rho\propto r^{-1.5}$.
\label{fig10}}
}
\end{center}
\begin{center}
{\leavevmode
\epsfxsize=13cm
\epsfbox{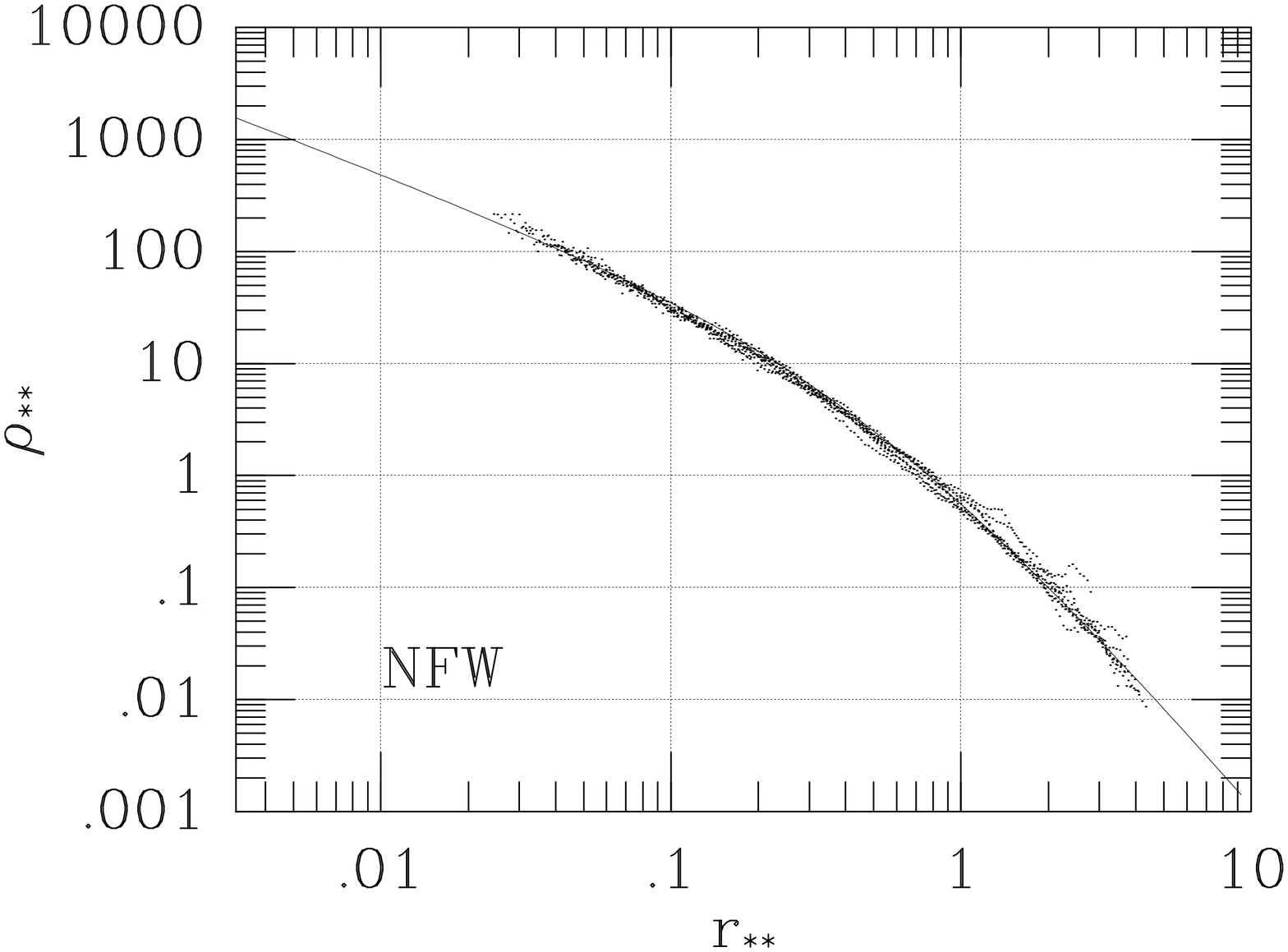}
\caption{
Fitting of the density structure by the profile proposed by Navarro, Frenk, 
White (1996,1997). 
\label{fig11}
}
}
\end{center}
\end{figure}

\begin{figure}
\begin{center}
{\leavevmode
\epsfxsize=17cm
\epsfbox{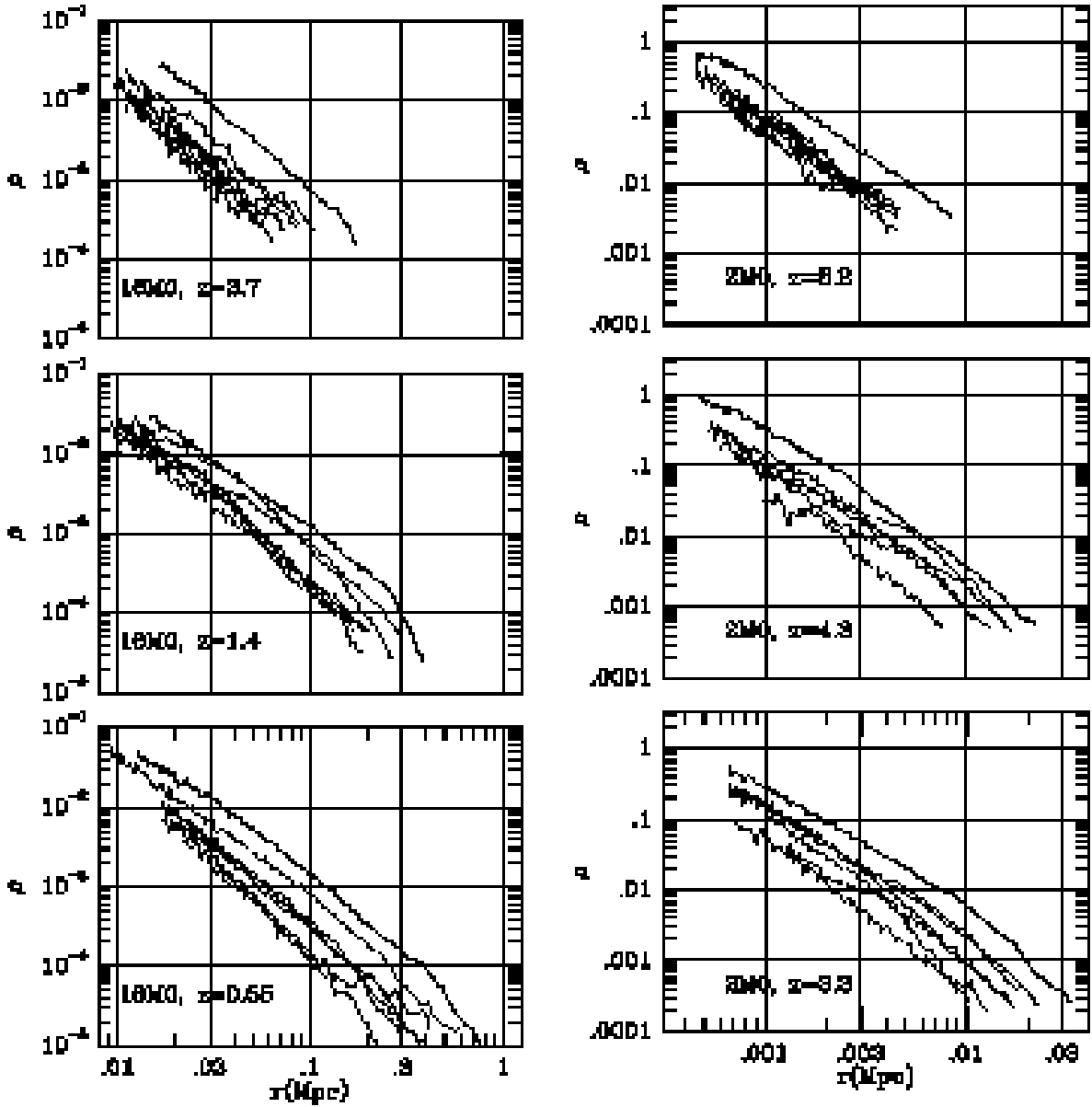}
\caption{Density profiles of smaller halos which are going to merge to
the larger halo.  The thick curve indicates the larger halo. 
\label{fig12}}
}
\end{center}
\end{figure}

In Figure \ref{fig7}, we can clearly see that the cusp grows outward
without changing the inner part.  In Figure \ref{fig12} we show density
profiles of halos which will merge to the largest halo, for Runs 16M0
and 2M0.  We plot 6 halos with more than 1000 particles which are
nearest from the potential minimum, together with the largest halo.  It
is clear that the central halo has the highest density. 

The reason why larger halos have higher density can be understood as
follows.  Let us consider the peaks of the fluctuation whose
characteristic scale is $\lambda$.  If the total density field is only
composed of the fluctuation whose scale is $\lambda$, the peaks will
collapse to halos with similar density almost simultaneously.  Actually,
there are contributions from fluctuations whose scale is larger than
$\lambda$, too.  A peak in high-density background would collapse to a
halo with higher density than peaks in low-density background, simply
because of the difference in the background density.  Later, the
"background", which is just a density peak of longer wavelength, would
collapse.  During this collapse, however, the high-density peak
collapsed earlier is not affected.  Therefore, larger halo tend to have
higher central density than smaller halos. 

\begin{figure}
\begin{center}
{\leavevmode
\epsfxsize=14cm
\epsfbox{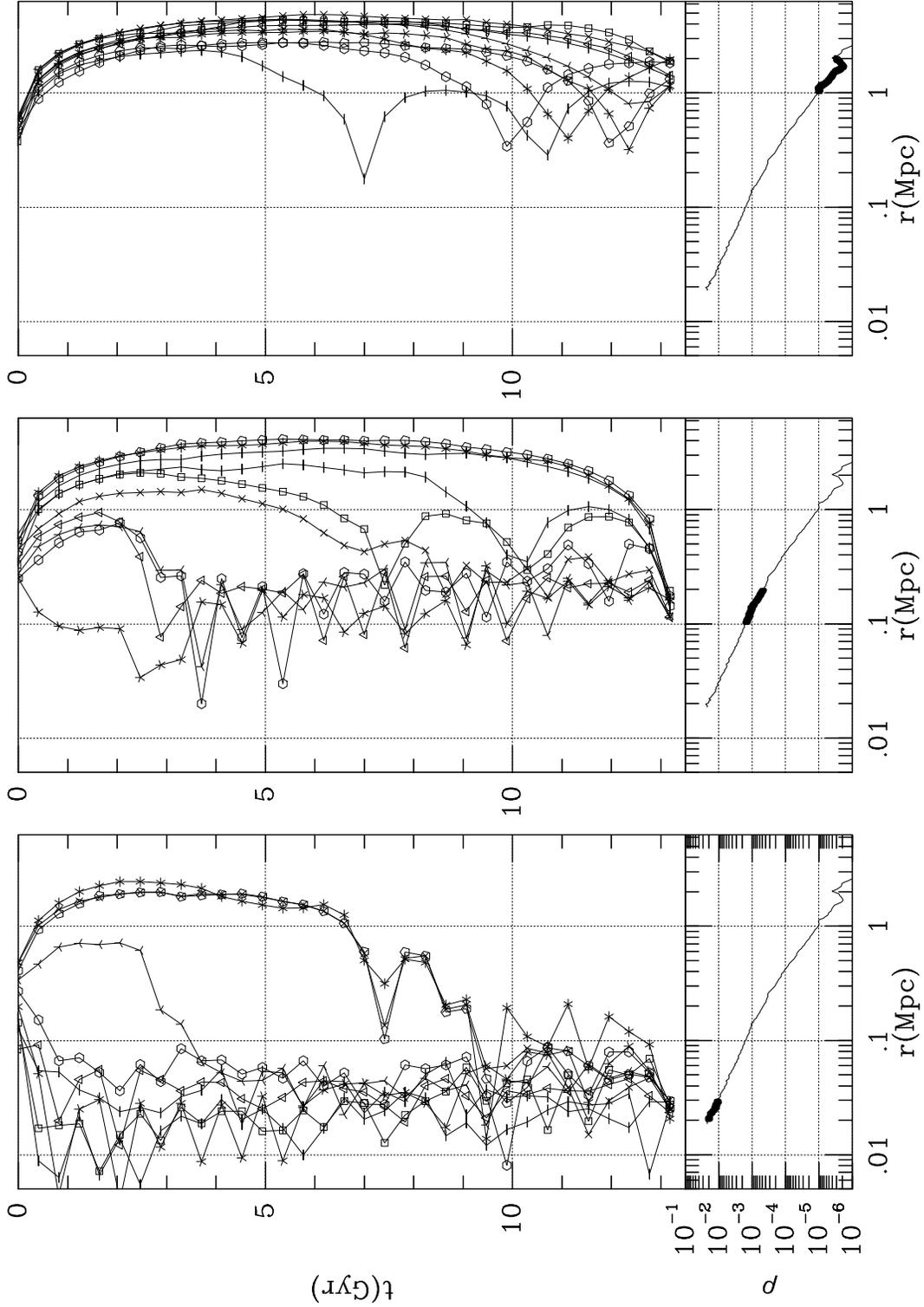}
\caption{One-dimensional trajectories of randomly selected 10 particles
for Run 16M0.  The distance from the center of halo is plotted as a
function of time, together with the density profile at $z_{\rm end}$.  
The region indicated by thick curve 
in density profile indicates where the selected particles exists at $z_{\rm end}$. 
\label{fig14}}
}
\end{center}
\end{figure}

In Figure \ref{fig14} we show one-dimensional trajectory of the
particles for Run 16M0.  We randomly select 10 particles from the
particles whose distances from the center of the halo at the end of the
simulation are 0.02-0.03, 0.1-0.2, and 1-2 Mpc Figure \ref{fig14} shows
that a large fraction of particles in the inner region settles there
early, and those in the outer region tend to fall later.  In other
words, Figure \ref{fig14} shows that the formation process discussed in
the above actually takes place. 

\label{seccusp}

The cusp with the slope of $-1.5$ seems to be a "fixed point" or
a "convergence point" for the growth of the halo through accretion of
diffuse and small halos.  Once the cusp with the slope of $-1.5$
forms, the density in the $r^{-1.5}$ cusp remains unchanged and the
disrupted matter is accreted outside the $r^{-1.5}$ cusp, which is
clearly shown in Figure \ref{fig7}. 

Moreover, the power index of $-1.5$ seems to be a universal feature
independent of the form of initial power spectrum.  The high-resolution
simulations presented in this paper and those by Moore et al.  (1999)
show that the power of the cusp is $-1.5$, independent of the mass
scale.  A preliminary result of our another simulation from the initial
power spectrum of $P(k) \propto k^{-1.7}$, which is shallower than that
at cluster scale for standard CDM model, also show that the power of the
cusp is around $-1.5$. 

Currently, we do not have a clear explanation why the slope of the cusp
is $-1.5$ when it forms through the accretion of disrupted small halos. 
We will discuss this topic more comprehensively elsewhere. 

\subsection{Origin of The Outer Profile}

\begin{figure}
\begin{center}
{\leavevmode
\epsfxsize=15cm
\epsfbox{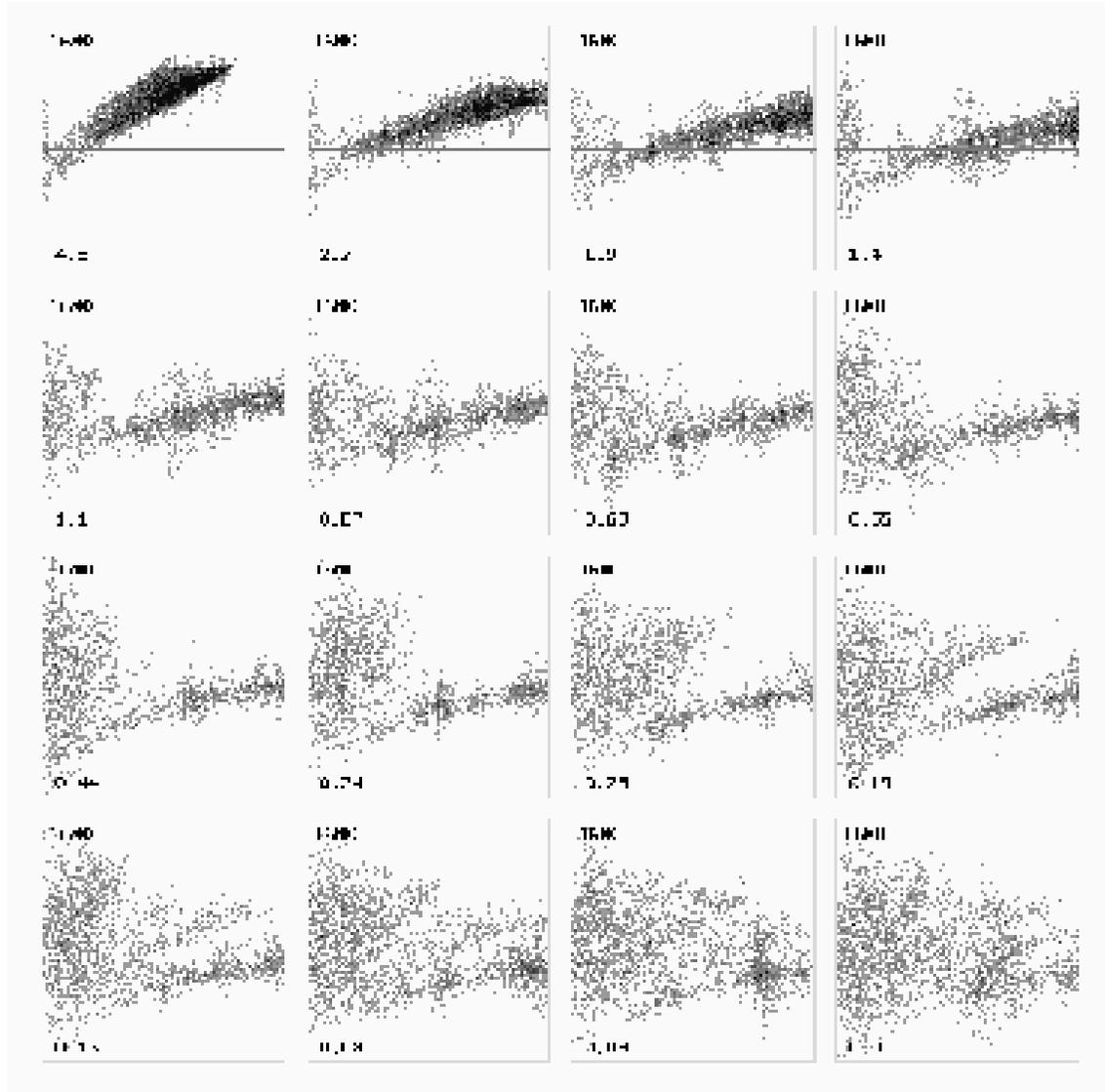}
\caption{Distribution of particle in $r-v_{\rm r}$ plane
at 16 different redshifts for $N$-body simulation Run 16M0. 
The number at lower-left corner of each panel indicates the redshift. 
The width of each panel is equal to $2r_{200}$ at $z_{\rm end}$. 
The range of radial velocity is from $-2031$ km/s to $2031$ km/s.
 \label{fig15}}
}
\end{center}
\end{figure}

\begin{figure}
\begin{center}
{\leavevmode
\epsfxsize=13cm
\epsfbox{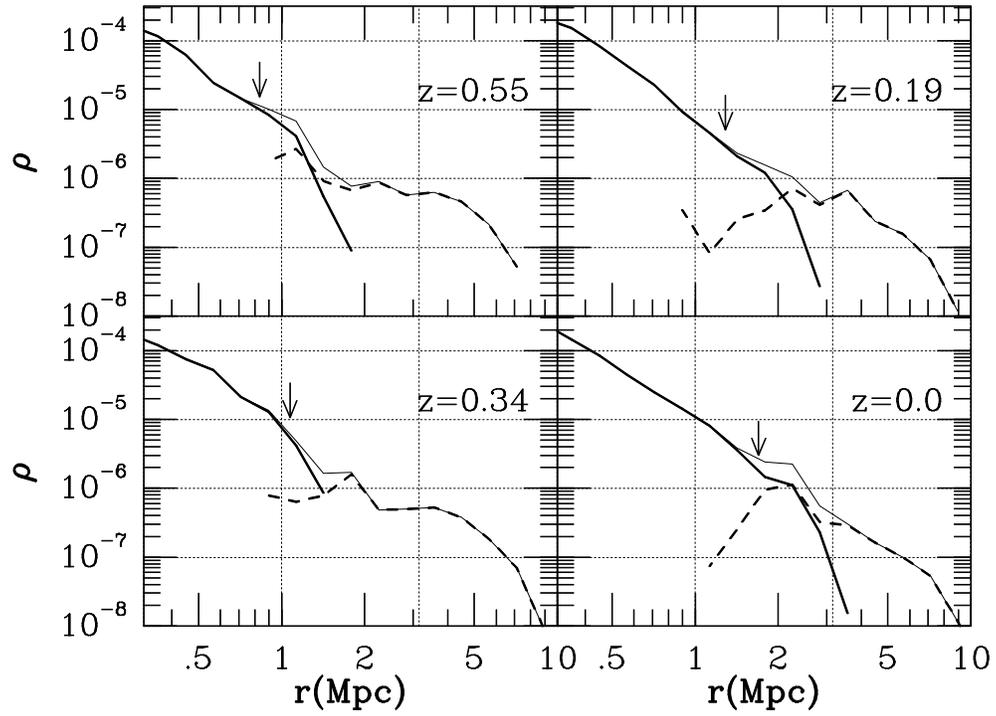}
\caption{
Density profile of the halo for Run 16M0 as a superposition of two
components.  The dashed curves indicate that for the infalling matter
and the thick curves indicate the scattered matter.  The thin curve is
total profile. 
\label{fig16}
}
}
\end{center}
\end{figure}

Figure \ref{fig15} shows the distribution of particles on the
$r$-$v_{\rm r}$ plane, where $r$ is distance from the center and $v_{\rm
r}$ is the radial velocity, at 16 different redshifts for Run 16M0.  We
can see that the outer region consists of two component.  The first
component is infalling matters which is visible as thick stream of
particles in the right-lower region of each panels.  The vertical
spreads visible in this stream are infalling smaller halos.  The second
component is the more scattered particles with nearly zero average
velocity.  As one can see from the time evolution, these stars gained
energy in the central region when small halos accreted on the central
halo.  In Figure \ref{fig16} we show density profiles of scattered and
infalling particles separately.  We separated two components by defining
appropriate boundary in Figure \ref{fig15}.  At around $r_{200}$,
contribution of two components to the total profile are of the same
order. 

The profile in the outer region exhibits large fluctuations.  The
merging events occur intermittently, and the amount of scattered matter
depends on earlier merging events.  Nevertheless, the density profile
fits the profile which is asymptotically proportional to $r^{-3}$, as
shown in the previous section. 

\begin{figure}
\begin{center}
{\leavevmode
\epsfxsize=13cm
\epsfbox{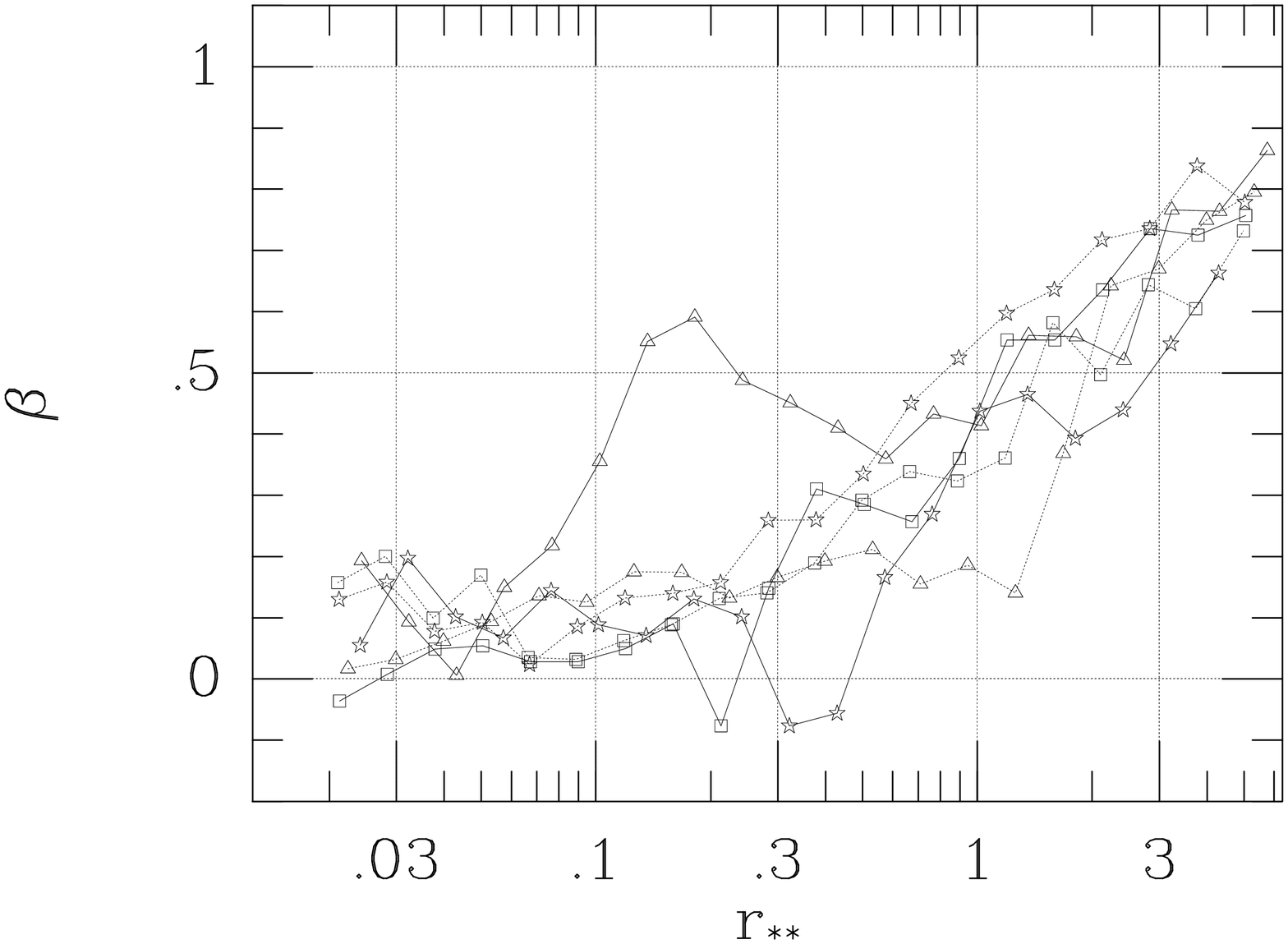}
\caption{
Anisotropy parameter, $\beta=1-v^2_{\theta}/v_{\rm r}^2$ as a function
of radius at $z=z_{\rm end}$ for Runs 16M\{0,1,2\}(solid curve) and
2M\{0,1,2\} (dashed curves).  The star, square, and triangle symbol
indicate $\beta$ for \{16,2\}M0, \{16,2\}M1, and \{16,2\}M2,
respectively.  \label{fig17}
}
}
\end{center}
\end{figure}

Consequently, in the outer region orbits of particles shows strong
radial anisotropy.  Figure \ref{fig17} shows anisotropy in velocity
distribution of the profile together with simulation results for Run
16M\{0,1,2\} and 2M\{0,1,2\}.  The anisotropy are expressed by the
anisotropy parameter $\beta$, defined as 
\begin{equation}
\beta= 1- {\langle v_{\theta}^2 \rangle  \over \langle v^2_{\rm r} \rangle},
\end{equation}
where $\langle v_{\theta}^2 \rangle$ and $\langle v^2_{\rm r} \rangle$
are mean tangential and radial velocity dispersion.  In this definition,
$\beta=0$ means the velocity distribution is isotropic and $\beta=1$
means completely radial. 

\section{Conclusion}
 
We performed $N$-body simulations of dark matter halo formation in the
standard CDM model.  We simulates 12 halos whose mass range is
$6.6\times 10^{11}M_{\odot}$ to $8.0\times 10^{14}M_{\odot}$.  We
introduced the accuracy criteria to guarantee that numerical artifact
due to the two-body relaxation and the time integration do not affect
the result, and obtained the density profile which is free from
numerical artifact down to the radii $(0.01-0.02)r_{200}$. 

Our main conclusions are: 

\begin{itemize}

\item[(1)] 
In all runs, the final halos have density cusps proportional to
$r^{-1.5}$. 

\item[(2)] 
The density profile evolves self-similarly.  

\item[(3)] 
The density profile is universal,
independent of the halo mass, initial random realization of density
fluctuation and the redshift.  The density structure is in good
agreement with the profile proposed by Moore et al.  (1999). 

\item[(4)] 
We found that the central cusp grows through the disruption and
accretion process of diffuse smaller halos.  The slope of the central
cusp seems to be a fixed point for the growth of the halo through
accretion of tidally disrupted matter. 

\end{itemize}

\section{Discussion}

Here we discuss the relation between our results and those of the
previous studies. 

We obtained steeper inner cusp than that obtained by NFW, which was
already found in high-resolution simulations (FM97, Moore et al 1999,
Ghigna et al.2000, Jing and Suto 2000).  The reason for this
disagreement is that in low-resolution simulations the central cusp is
smoothed out by the two-body relaxation.  If the cusp is shallower than
$-2$, the velocity dispersion decreases inward.  The energy flows inward
due to the two-body relaxation and the central region expands, which is
called the gravothermal expansion (Hachisu et al.  1978, Quinlan 1996,
Heggie, Inagaki, McMillan 1994, Endo, Fukushige, Makino 1997). 
Therefore, the density in the cusp decreases and the cusp becomes
shallower.  Using the relations $t_{\rm rel} \sim v^3/(\rho m)$,
$\rho\sim r^{-1.5}$, and $v\sim r^{0.25}$ and our simulation results, we
can estimate the lower limit of the radius where the structure is free
from the two-body relaxation effect as $\sim 0.01
r_{200}(N/10^6)^{-0.44}(\rho_0/2.7\times 10^{-4}M_{\odot}/{\rm
pc}^3)^{0.44}(\sigma_0/1300 {\rm km/s})^{-1.33}$ for a cluster-sized
halo at the present epoch, where $\sigma_0$ is velocity dispersion at
the scale radius $r_0$.  For simulations with $N=10^4$ and $10^5$ within
$r_{200}$, the limits are estimated as $\sim 0.08r_{200}$ and $\sim
0.03r_{200}$, respectively.  Therefore, simulations with $\sim 20$k
particles the central cusp would be become significantly shallower due
to relaxation. 

We could not reproduce the dependence of the slope observed by Jing and
Suto (2000).  The difference could be also due to the smoothing by
two-body relaxation in their cluster-sized halos.  In this paper, we
show that the density profile within $\sim 0.01r_{200}$ smoothed by the
two-body relaxation.  The density at $0.01r_{200}$ and the mass
resolution in their cluster-sized halo are similar to those in ours. 
The density profile in their simulations within $0.01r_{200}$, at which
the profile begins to depart from $r^{-1.5}$ inward, could be affected
by the two-body relaxation. 

Our result is in good agreement with results of simulations by Moore et
al.  (1999), in which the tidal field was included.  This agreement
suggests that the neglect of the tidal field in our present study and in
FM97 hardly affects the density profile.  The formation mechanism we
discussed in this paper also suggests that the tidal field due to the
mass outside the simulation sphere is not crucial. 

Moore et al.  (1999) argued that the merging process is not related with
the structure.  They simulated the halo formation using a power spectrum
with a cutoff to suppress merging event in smaller scale and showed the
profile does not change.  However, in this simulation, several merging
event took place since the cutoff wave length is rather short. 
Therefore, their conclusion that merging is not important is not really
supported by their simulation.  According to our explanation, several
merging events where the central large halo swallows smaller halos
determines the structure of the halo.  Such events did took place in
Moore et al.'s simulations. 

As discussed in the Introduction, there are a lot of analytical and
semi-analytical studies to explain the density structure.  However, no
study succeeded to explain the universality satisfactory.  This is
because none of them is based on the formation and growth process of the
halo as discussed in this paper.  Syer and White (1998) discussed that
the cusp is a convergence point of disrupting and sinking of satellite,
and that the convergence slope depends on the initial power spectrum. 
In their model, some of smaller halos were assumed to sink down to the
center of the larger halo, when two halos merged.  However, as we
discussed, smaller halos are always disrupted and never sink down to the
center, because the smaller halo is always less dense.  Evans and Collet
(1997) argued the cusp is a steady-state solution of Fokker-Plank
equation.  The solution is derived by assuming that many small clumps
within a large halo evolve by the two-body relaxation.  In our
simulations, there are no such small clumps, since they are disrupted
before they reaches the center. 

\acknowledgements

We are grateful to Atsushi Kawai, for his help in preparing the hardware
and software environment of the GRAPE-5 system, and to Yasushi Suto and
Yoko Funato for many helpful discussions.  To generate initial
condition, we used the COSMICS package developed by Edmund Bertschinger,
to whom we express our thanks.  A part of numerical computations was
carried out on the GRAPE system at ADAC (the Astronomical Data Analysis
Center) of the National Astronomical Observatory, Japan.  This research
was partially supported by the Research for the Future Program of Japan
Society for the Promotion of Science, grant no.  JSPS-RFTP 97P01102.

\clearpage

\end{document}